**Review:**

# Band-like Carriers in a Soft, Anharmonic Lattice: Lead-Halide Perovskites

Young Mi Lee[1,†], Inhee Maeng[2,†], Jinwoo Park[3], Seung-Jae Oh[2], and Min-Cherl Jung[4,*]

[1]*Beamline Department, Pohang Accelerator Laboratory, POSTECH, Pohang 37673, Republic of Korea*

[2]*YUHS-KRIBB Medical Convergence Research Institute, College of Medicine, Yonsei University, Seoul 03722, Republic of Korea*

[3]*Department of Physics, University of Seoul, Seoul 02504, Republic of Korea*

[4]*Global Institute of Future Technology, Shanghai Jiao Tong University, Shanghai 200240, China*

[†]*These authors contributed equally to this work.*

E-mail: mincherl.jung@sjtu.edu.cn

## Abstract

Lead-halide perovskites $APbX_3$ present a striking dichotomy: they exhibit band-like electronic transport despite their exceptionally soft and anharmonic lattices. Optically and electrically, they resemble conventional direct-gap semiconductors, exhibiting light carrier masses and steep absorption onsets, whereas their lattices display liquid-like dynamics, including overdamped octahedral motions, quasielastic Raman central peaks, and exceptionally low thermal conductivities. Solution-processed films nevertheless sustain micrometre-scale carrier diffusion at defect densities that would severely suppress transport in conventional semiconductors. Here, we argue that these apparently disparate properties emerge from a common microscopic framework: a soft, strongly anharmonic, and polar $[PbX_3]^-$ framework, strongly influenced by the Pb 6$s^2$ lone pair, which simultaneously shapes the antibonding orbital character of the band edges, the magnitude and multiple timescales of the dielectric response, and the slow relaxational dynamics that dress every charge carrier. We therefore invert the conventional order and develop the lattice before the electronic structure, because the nominally cubic phase is better viewed as a thermally fluctuating ensemble of locally symmetry-broken configurations rather than a single geometry. Within this framework, we discuss excitons, Fröhlich large polarons in the intermediate-coupling regime, carrier transport, defect tolerance, dimensional reduction in two-dimensional and nanocrystalline derivatives, and symmetry-breaking phenomena. We critically assess three contested issues — defect tolerance, ferroelectricity, and the interpretation of the $T^{-3/2}$ mobility law — and identify seven open questions together with the key measurements needed to resolve them.

# 1. Introduction

## 1.1. A material that resists classification

Few crystalline solids have been characterized so intensively and yet remain so difficult to place within the standard taxonomy of condensed matter, as the lead-halide perovskites $APbX_3$. On the basis of their optical and transport spectra they are unremarkable direct-gap semiconductors: a sharp absorption onset with an Urbach tail as steep as those of the best inorganic absorbers, a band-edge photoluminescence line of modest width, carrier effective masses of order one tenth of the free-electron mass[1], and mobilities[2] in the tens of $cm^2V^{-1}s^{-1}$. Every one of these observations invites the standard machinery: Bloch states, a parabolic effective-mass expansion about the band extrema, and weak scattering treated perturbatively.

On the evidence of their structural probes, they are something else entirely. Low-frequency Raman scattering in the room-temperature phases does not show the sharp, well-separated phonon lines of a conventional semiconductor; it shows a strong quasielastic *central peak* — intensity concentrated at zero frequency shift and decaying smoothly outward — which is the classic spectroscopic signature of an overdamped, relaxational degree of freedom rather than an oscillatory one[3] (Fig. 1). The distinction matters and is worth stating carefully. A damped harmonic mode of frequency $\omega_0$ and damping $\gamma$ produces a spectral peak *at* $\Omega_0$ so long as $\gamma < 2\,\Omega_0$; when the damping exceeds that bound the oscillation ceases to complete even a single period, the peak collapses onto zero frequency, and the coordinate no longer oscillates but *relaxes*, as in a liquid, obeying

$$d^2x/dt^2 + \gamma\,(dx/dt) + \Omega_0^2 x = 0 \qquad (1)$$

What the Raman experiments report is that a substantial part of the low-frequency lattice response of cubic $APbX_3$ sits on the wrong side of that bound. Momentum-resolved probes sharpen the picture: inelastic neutron and X-ray scattering on $CsPbBr_3$ resolve rods of diffuse intensity running along the edges of the cubic Brillouin zone, which in real space correspond to head-to-tail rotations of the $PbBr_6$ octahedra that are correlated within two-dimensional planes but essentially uncorrelated from plane to plane[4] (Fig. 2).

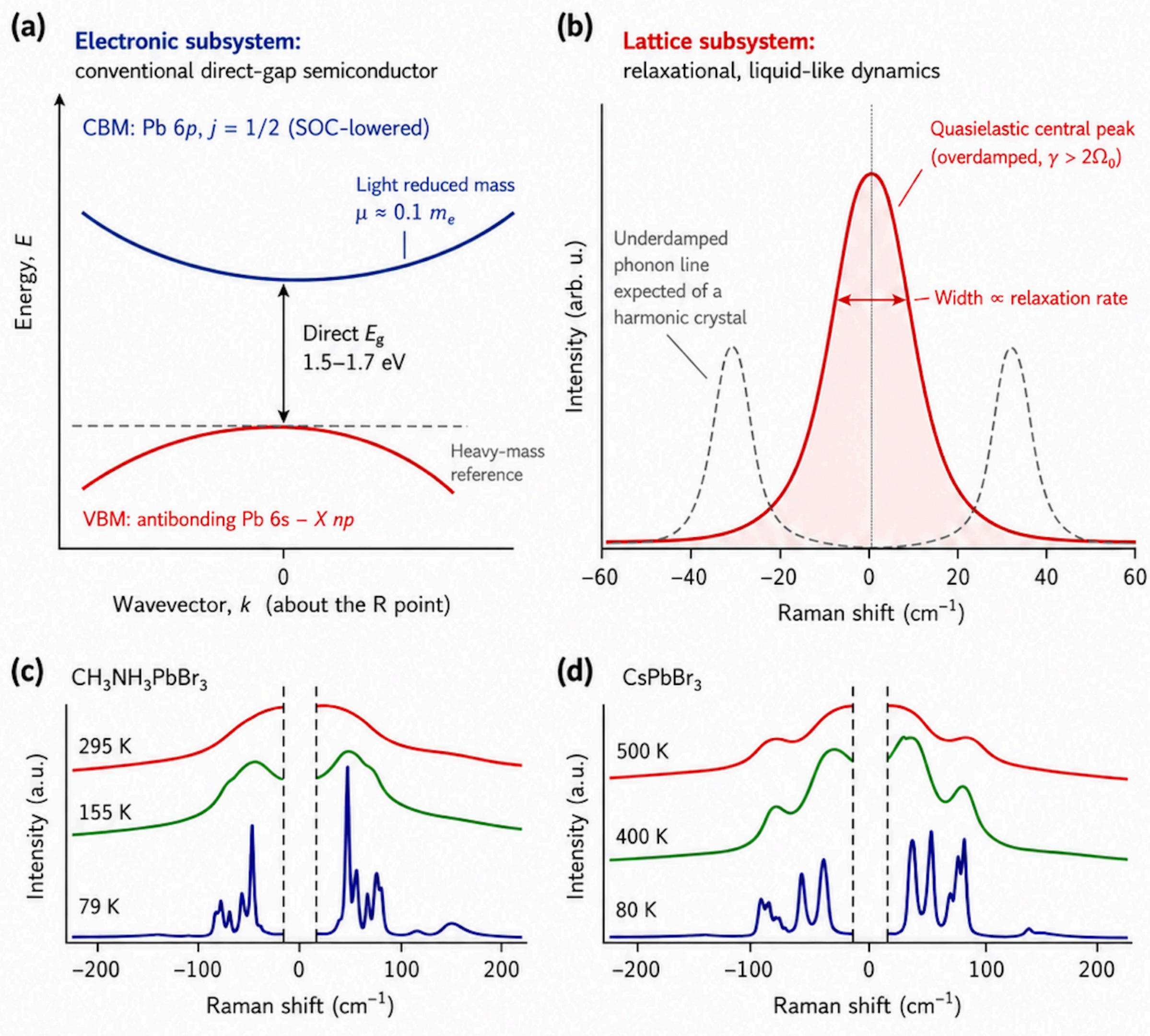

**Fig. 1 | The dual character of $APbX_3$ in schematic form. (a)** The electronic subsystem is that of a conventional direct-gap semiconductor: dispersive Bloch bands, a light reduced mass near 0.1 $m_e$, and a valence-band maximum of antibonding Pb 6*s* - X np character (Section 5). **(b)** The lattice subsystem, by contrast, exhibits the signature of relaxational rather than oscillatory dynamics in its high-temperature phases: instead of the underdamped phonon line expected of a harmonic crystal (dashed), low-frequency Raman spectra are dominated by a quasielliptical central peak (solid), indicating overdamped octahedral dynamics (Section 4). **Low-frequency Raman spectra of hybrid and inorganic lead-halide perovskite crystals**. The spectral region obscured by the notch-filter, between ±10 $cm^{-1}$ (marked by the vertical dashed lines), has been deleted. **(c)** $CH_3NH_3PbBr_3$ and **(d)** $CsPbBr_3$ in the orthorhombic phase (blue), tetragonal phase (green), and cubic phase (red), showing growth of central peak with temperature in both materials. Reprinted with permission from Yaffe et al., Phys. Rev. Lett. 118, 136001 (2017) https://doi.org/10.1103/PhysRevLett.118.136001. Copyright 2017 by the American Physical Society.

"Figure adapted from [4], reproduced with permission in the final published version"

**Fig. 2 | Diffuse scattering revealing network of rods from overdamped phonons. a–c**, Maps of diffuse-scattering intensity S(Q) for Q in the ($H$, $K$, $L$ = 0.5) reciprocal plane (cubic notation). K > 0 shows X-ray diffraction data from beamline 6-ID-D at the Advanced Photon Source, measured in the orthorhombic (333 K) **(a)**, tetragonal (373 K) **(b)** and cubic (433 K) **(c)** phases. K < 0 shows neutron data (energy integrated ±10 meV) from CNCS, measured in the orthorhombic (300 K) **(a)**, tetragonal (375 K) **(b)** and cubic (419 K) **(c)** phases. r.l.u., reciprocal lattice units. **d**, Three-dimensional rendering of the diffuse rod geometry (CNCS, −1 < E < 1 meV). **e,f,** Comparison of TDS simulation (K > 0) and CNCS data (K < 0) in the cubic phase (plane $L$ = 0.5) for low energy transfers (−2 < E < 2 meV) **(e)** and higher energy transfers (2 < E < 10 meV) **(f)**. Reproduced with permission from ref. 4, Springer Nature Ltd.

A material whose carriers propagate as Bloch states through a lattice whose instantaneous configuration is liquid-like is not, on the face of it, a consistent object. This tension is the subject of the present review, and it is not merely a matter of reconciling two sets of measurements. The band picture and the disordered-lattice picture invoke incompatible small parameters. The former presumes that the deviation of the ionic positions from a periodic reference structure is a weak perturbation; to be absorbed into a self-energy whose imaginary part gives a scattering rate small compared with the band energies. The latter reports instantaneous displacements — octahedral rotations of several degrees, halide excursions of tenths of an Å — comparable to the structural changes that separate entire phases and that shift the band gap by hundreds of meV. That both pictures should nonetheless yield quantitatively useful predictions is the central fact requiring explanation, and it is a fact about the *coupling* between the electronic and lattice subsystems rather than about either one separately.

## 1.2. The puzzle in its sharpest form

The puzzle admits a compact statement. Solution-processed lead-halide perovskite films are grown at or near room temperature from precursor solutions and, by any conventional standard of semiconductor metallurgy, are structurally poor: they are polycrystalline with sub-micrometre grains, chemically inhomogeneous at multiple length scales, and carry native point-defect densities orders of magnitude above those tolerated in silicon or gallium arsenide, whose usable perfection is purchased with kilokelvin processing and nine-nines purity. Nevertheless, ambipolar carrier diffusion lengths exceeding one micrometre were reported already in the earliest careful transport measurements on such films[6], and nonradiative lifetimes in good single crystals reach the microsecond scale. Frost[5] formulates this as the material's central enigma: how can a solid be so defective and still perform so well?

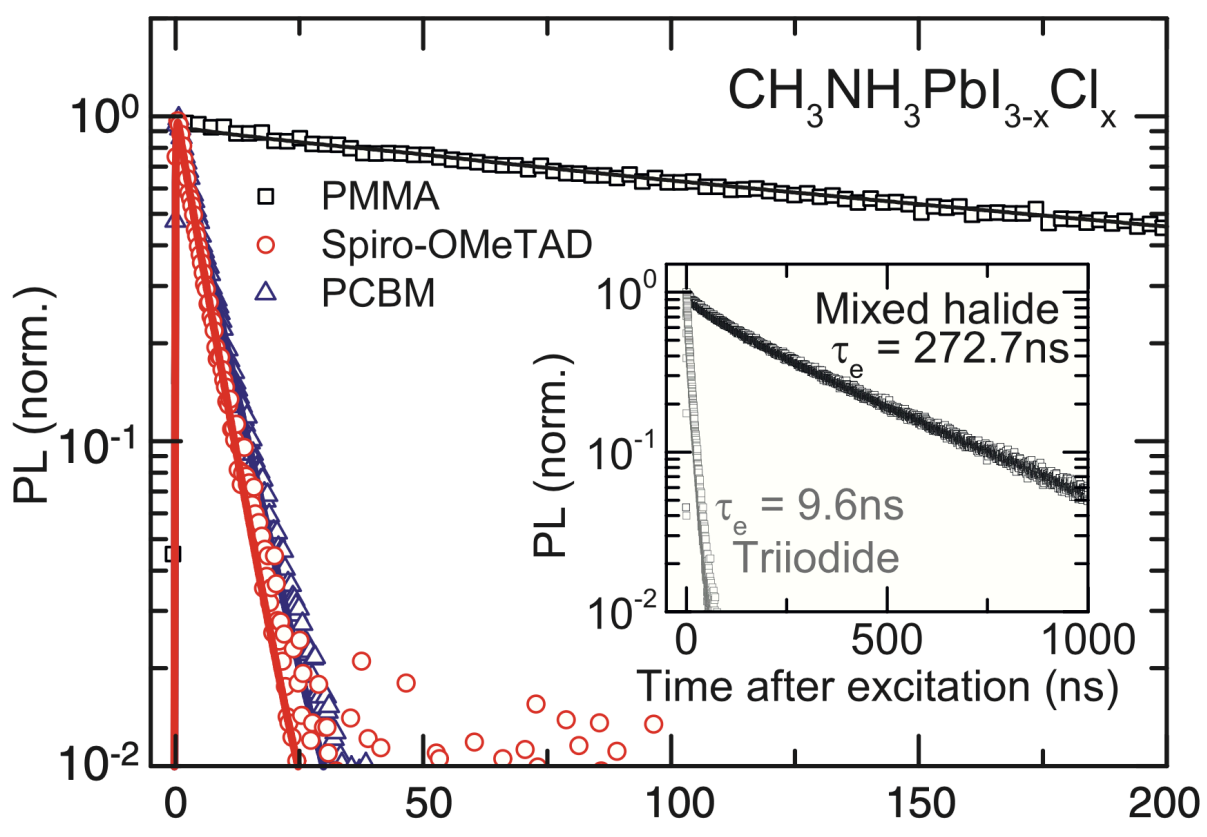

**Fig. 3 | Ambipolar carrier diffusion in a solution-processed mixed-halide perovskite film:** time-resolved photoluminescence quenching against a selective quenching layer, and the diffusion length extracted from the fit, exceeding one micrometre despite the film's modest structural quality[6]. Reproduced with permission from Stranks et al., Science 342, 341–344 (2013). 

Three broad classes of explanation have been advanced. We set them out here as distinct proposals, since that is how they entered the literature and how they are still, in places, debated; each requires a short elaboration, because each imports a definite piece of condensed-matter machinery whose applicability must later be examined.

The first is *electronic*, and concerns where defect levels sit. In a tetrahedral semiconductor such as GaAs the valence-band maximum is built from bonding combinations of anion p orbitals; breaking bonds — which is what a vacancy does — releases nonbonding states into the gap, where they act as deep traps and recombination centres. In $APbX_3$ the ordering is inverted: the valence-band maximum derives from the *antibonding* combination of the Pb 6s and halide np orbitals, pushed up by the lone-pair interaction, while the conduction-band minimum is predominantly Pb 6p, pulled down by spin-orbit coupling. Dangling-bond-like states associated with the low-formation-energy native defects then tend to appear within, or resonant with, the band continua rather than deep in the gap — the essential content of the defect calculations of Yin, Shi, and Yan[7]. On this view the material is not free of defects; it is indifferent to the ones it has. Whether this "defect tolerance" survives more sophisticated treatments of charge capture is contested[8,9], and Section 9 treats the dispute at length.

The second is *dielectric and polaronic*, and concerns what the carrier is. The lattice is strongly polar: the Born effective charges substantially exceed the formal ionic charges, and the low-lying longitudinal optical phonons — at 11.5 and 15.3 meV in $MAPbI_3$ and $MAPbBr_3$ respectively — couple to carriers through the long-range Fröhlich interaction with characteristic coupling energies of order 40-60 meV [13]. In a polar lattice a slow charge does not remain a bare band electron: it polarizes its surroundings, and electron plus polarization cloud together form a Fröhlich *large polaron* — "large" meaning that the dressed object extends over many unit cells, so that the band picture survives with a renormalized mass, in contrast to the small polaron whose carrier is self-trapped on a single site and moves by hopping. The proposal of Zhu and Podzorov[10], developed experimentally by Zhu *et al.*[11] and Miyata *et al.*[12], is that this polarization cloud does double duty: it renormalizes the carrier, and it screens the carrier from charged defects and from further phonon scattering. The mobility one measures is then the mobility of the dressed object, and the benignity of defects is partly a property of the dressing rather than of the defect chemistry.

The third is *dynamical* and concerns the timescale of the dressing. In a conventional polar semiconductor, the lattice polarization follows the carrier essentially instantaneously on the scale of the scattering time, and the static and dynamic descriptions coincide. Here the polarization that dresses the carrier is anharmonic and slow — the relaxational dynamics of Section 1.1 above — evolving on picosecond timescales comparable to, or slower than, the carrier scattering time itself. Miyata, Atallah, and Zhu[14] capture this regime as a "crystal-liquid duality": the material behaves as a crystal for the electrons and as a liquid for the phonons, in the same sense as the phonon-glass electron-crystal concept familiar from thermoelectrics, where it is engineered deliberately and here arises intrinsically.

These three proposals may represent different, strongly coupled consequences of the soft, anharmonic, polar character of the $[PbX_3]^-$ framework, which influences the orbital character of the band edges, the magnitude of the dielectric response, and the timescale of that response. We regard this coupling as a useful interpretation of the present evidence, and it is the hypothesis that organizes this review; Figure 4 displays it as the causal structure on which the sections are arranged. We therefore treat this as a hypothesis rather than an established conclusion. The evidence required to defend it is developed in Sections 4 through 6, and the argument is made explicitly in Section 7, where the polaron and the anharmonic lattice are treated together; the evidence for this hypothesis is developed in Section 7.5 as the central section of this review. Where the unification fails — and Section 9 will argue that the defect problem is not yet reducible to it — we say so.

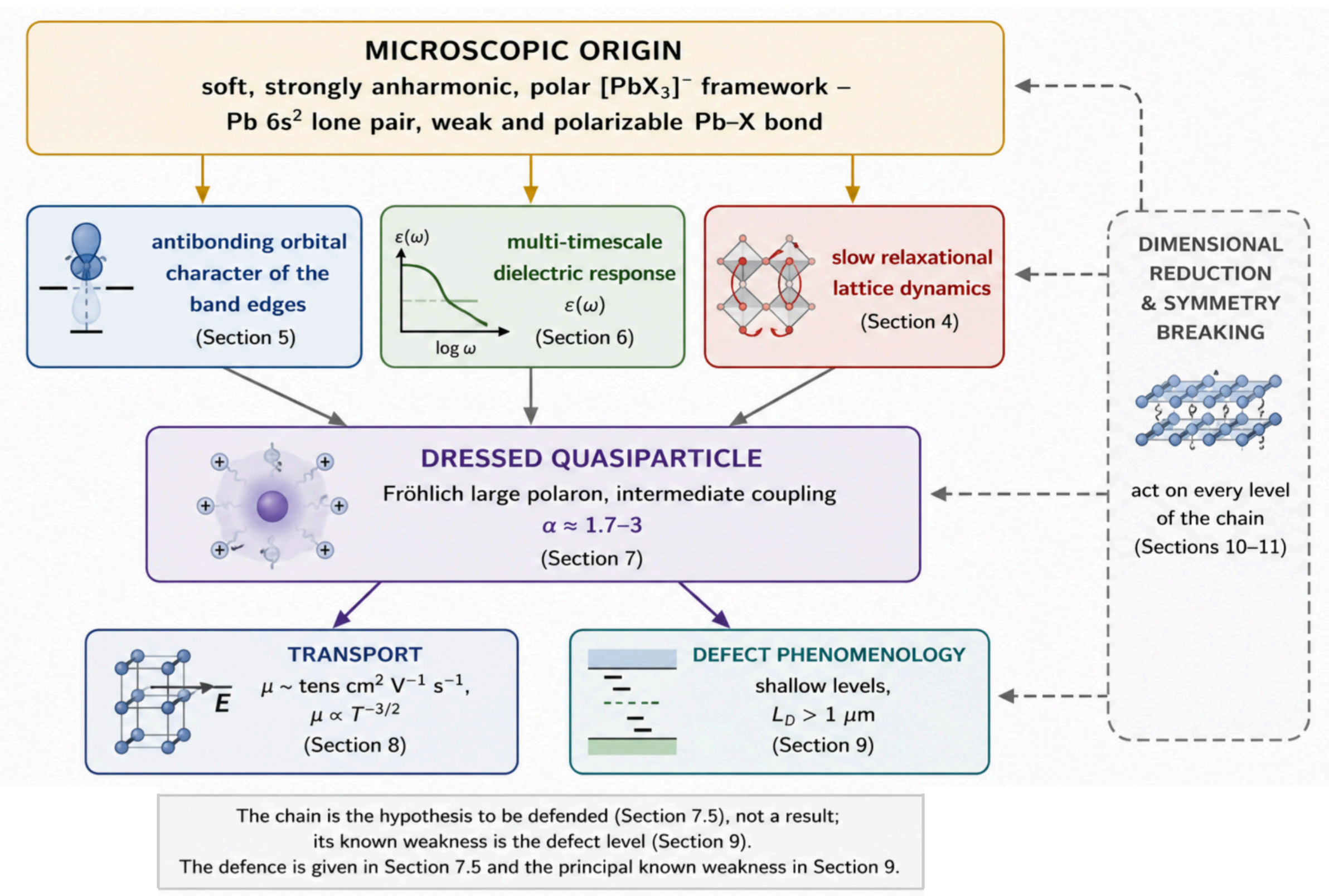

**Fig. 4 | The organizing hypothesis of this review, displayed as a causal chain.** The soft, anharmonic and polar $[PbX_3]^-$ framework, whose character is influenced by the Pb $6s^2$ lone pair and the weak, polarizable Pb–X bond, feeds three intermediate descriptions — the orbital character of the band edges, the multi-timescale dielectric response and the slow relaxational lattice dynamics — which together shape the dressed quasiparticles of Section 7 and the observable transport and defect phenomenology. Dashed line: the modifications introduced by dimensional reduction and symmetry breaking (Sections 10–11) modify the relative importance of these ingredients. The status of this diagram is that of a hypothesis to be tested, not a result; the argument is developed in Section 7.5 and its principal limitation is discussed in Section 9.

## 1.3. Why a condensed-matter treatment, and why now

The literature on these materials is dominated, by volume, by device and chemistry reports. This is neither surprising nor regrettable — the photovoltaic performance is what drew attention to the system — but it has had two effects that a condensed-matter review is well placed to remedy.

First, phenomenological explanations have propagated faster than they have been tested. Terms such as "defect tolerance," "hot-carrier bottleneck," and "ferroelectric enhancement" entered the field as plausible mechanisms and were subsequently repeated as established fact, often shorn of the qualifications their originators attached. Egger *et al.*[15] mounted the first systematic scrutiny of which properties of these materials are in fact anomalous when measured against an appropriate reference semiconductor — asking, property by property, whether the halide perovskite value lies outside the range spanned by GaAs, CdTe, and the chalcopyrites once the smaller gap and softer lattice are accounted for — and concluded that a substantial fraction of the claimed anomalies are either quantitatively modest or not yet demonstrated. Their essay is, in a sense, the point of departure for the present review, and we adopt its posture throughout: a claim of anomaly must be defended against a control. Where the all-inorganic compound reproduces the behavior of the hybrid, the organic cation is not the cause; where a soft polar semiconductor of ordinary chemistry would behave identically, no exotic mechanism should be invoked.

Second, the material has by now generated a body of measurement that is of intrinsic interest to condensed-matter physics independent of any application. Momentum-resolved inelastic scattering has exposed a form of correlated, low-dimensional structural fluctuation in a three-dimensional crystal with few clean analogues[4]. High-field magneto-absorption to 150 T has furnished exciton reduced masses to three-digit precision and settled a binding-energy controversy that conventional spectroscopy had left spanning an order of magnitude[1]. The system provides an unusually clean experimental realization of intermediate-coupling Fröhlich polaron physics — dimensionless coupling constants of order 1-3, between the weak-coupling perturbative regime and the strong-coupling self-trapped one — a subject that has been theoretically mature since Feynman but experimentally underdetermined for want of materials that sit in the interesting regime[16]. Strong spin-orbit coupling on a heavy p-block cation in a nominally centrosymmetric lattice raises the question of Rashba splitting generated by dynamic, rather than static, inversion-symmetry breaking[17]. None of these questions is about solar cells.

Third, this review sits deliberately alongside several excellent existing surveys rather than in place of them, and it is worth saying plainly what it adds. Egger *et al.*[15] audit which claimed anomalies of the material survive comparison against a conventional reference semiconductor, a discipline this review adopts throughout but does not attempt to redo at their level of exhaustiveness. Katan, Mercier, and Even[18] and Baranowski and Plochocka[62] treat the two-dimensional structural chemistry and the excitonic literature, respectively, in a depth that Sections 7 and 10 here only summarize and build on. Franchini *et al.*[16] situate the polaron physics within the broader materials class of polaronic semiconductors, and Mosquera-Lois *et al.*[9] audit the defect-tolerance mechanisms specifically, a synthesis Section 9 leans on rather than repeats. What we believe is new here is the architecture rather than any single fact within it: no prior treatment we are aware of builds the electronic structure explicitly as a *consequence* of the lattice rather than the reverse, states the unification of the three explanatory traditions of Section 1.2 as an explicit and falsifiable hypothesis with a dedicated defense and delimitation (Section 7.5), or closes with a numbered, resolvable research programmed (Section 12) in place of an unranked list of open questions. A reader seeking exhaustive coverage of any single sub-literature is better served by the specialist reviews cited above, read alongside this one; the contribution attempted here is the connective structure between them.

## 1.4. Scope

We restrict attention to lead-based systems: three-dimensional $APbX_3$ with A = MA, FA, or Cs and X = Cl, Br, I, together with the two-dimensional Ruddlesden-Popper and Dion-Jacobson derivatives obtained by slicing the same $[PbX_3]^-$ framework along low-index planes and intercalating organic spacers[18]. The restriction is a matter of physics rather than convenience. The $Pb^{2+}$ $6s^2$ lone pair, its stereochemical activity, and the relativistic magnitude of spin-orbit coupling on lead are precisely the ingredients that generate the phenomenology under discussion; substituting Sn or Bi changes the problem rather than extending it — the Sn $5s$ lone pair is more strongly expressed and drives different structural chemistry, while the double perovskites trade the direct gap for parity-forbidden or indirect transitions. Lead-free compounds therefore appear here only as controls, in the specific role of isolating which observations require the lone pair, and which do not. For the same reason the all-inorganic Cs compounds are *inside* our scope: they are the indispensable control for every claim about the role of the molecular cation, and the running comparison MA versus Cs is one of the review's principal analytical tools.

Applications are compressed into the historical narrative of Section 2 and are otherwise invoked only where a device measurement constitutes a physical measurement — an electroluminescence quantum yield as a probe of nonradiative channels, an open-circuit voltage as a probe of quasi-Fermi-level splitting. Photovoltaic efficiency tables, deposition chemistry, encapsulation, and stability engineering lie outside the scope of this review.

## 1.5. Organization, and a note on the ordering

Sections 3 and 4 establish the lattice: crystal structure, symmetry, tilt systems, and the phase-transition sequence in Section 3; then the dynamics — soft modes, anharmonicity, overdamping, and the distinction between the average and the instantaneous structure — in Section 4. Section 5 constructs the electronic structure on that foundation, including spin-orbit coupling, relativistic quasiparticle corrections, and the polymorphous description of the nominally cubic phase. Section 6 treats the dielectric response that connects the two subsystems. Section 7 assembles the quasiparticles — excitons, polarons, and their hybrids — and it is there that the unifying hypothesis stated in Section 1.2 is examined. Sections 8 and 9 draw out the consequences for transport and for defect behaviour. Section 10 reduces the dimensionality; Section 11 addresses broken symmetry and emergent phenomena; Section 12 collects what remains genuinely open. Three appendices summarize the theoretical machinery that recurs throughout: relativistic density-functional and GW-BSE practice (Appendix A), molecular dynamics on anharmonic energy landscapes (Appendix B), and the Fröhlich-Feynman polaron formalism (Appendix C).

The reader accustomed to semiconductor reviews will notice that this ordering inverts the usual one, in which the band structure is presented first, and electron-phonon coupling appears later as a correction. The inversion is deliberate, and since it determines the shape of the entire review it deserves a paragraph of justification. In silicon, the harmonic lattice is a background against which the electronic structure is defined: the atoms sit at their equilibrium sites to within a few per cent of the bond length, and anharmonicity enters as a small correction to thermal expansion and phonon linewidths. In $APbX_3$ the lattice is not a background. The instantaneous band gap fluctuates by an appreciable fraction of its mean value on the timescale of a few hundred femtoseconds[19,20]; the "cubic" phase, examined locally, is a distribution of symmetry-broken configurations whose *average* is cubic, so that the static band structure is a configurational average rather than a property of any equilibrium geometry[21]; and quantities as basic as the sign of the temperature coefficient of the gap come out wrong unless the anharmonic motion is included. Presenting the electronic structure first would therefore require us to unlearn it immediately. We build the lattice first and derive the electrons from it.

Throughout, each section is organized to separate what is measured from what is inferred. We first state the experimental facts; then the standard condensed-matter framework onto which they map; then the point at which that framework fails; and finally, the question that remains open. Where the literature is divided, we say so, and where we take a position, we mark it as ours.

# 2. Historical Development

The history of this material is unusual in that its physics was largely established before its application was conceived, then forgotten, then rediscovered under the pressure of a technological result. We recount it here not for antiquarian interest but because the sequence explains a persistent feature of the modern literature:

several questions now treated as open were addressed, with different instruments and different vocabulary, in work of the 1980s and 1990s, and at least one major controversy of the 2010s was settled by returning to a technique of 1994 with better apparatus. Figure 5 summarizes the chronology.

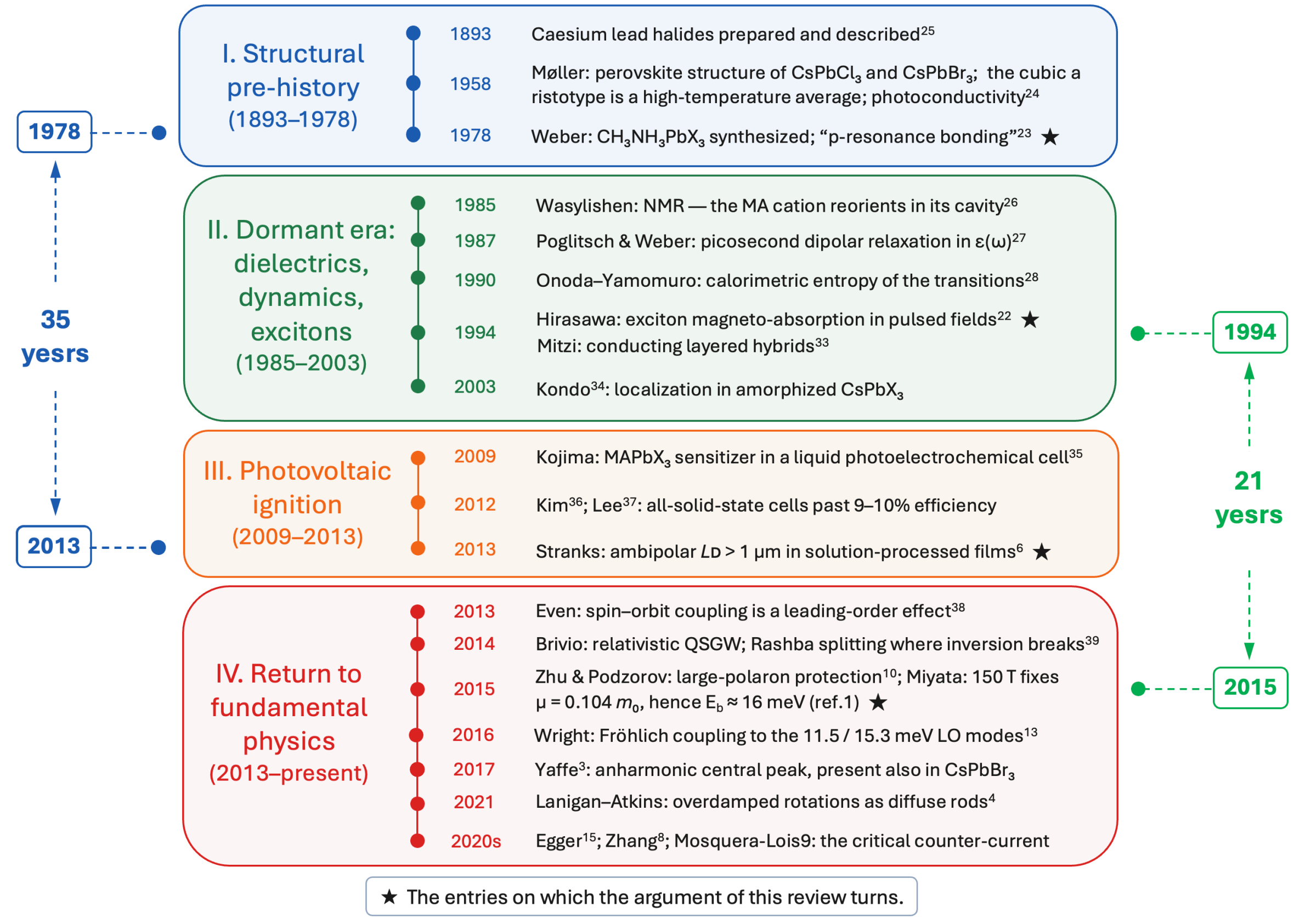

**Fig. 5 | Chronology of the field, divided into the four eras used in the text**: structural pre-history (1893-1978), the dormant era of dielectric, dynamic, and excitonic physics (1985-2003), photovoltaic ignition (2009-2013), and the return to fundamental physics (2013-present). Note the four-decade gap separating the synthesis of the hybrid compounds[23] from the transport measurement[6] that converted them into a physics problem, and the position of the exciton magneto-optics of Hirasawa *et al.*[22] a full two decades before the same technique, at higher field, closed the binding-energy controversy.

## 2.1. Structural pre-history, 1893-1978

Cesium lead halides were prepared and described in the closing years of the nineteenth century[25], but their structural identification belongs to Møller[24], who established that $CsPbCl_3$ and $CsPbBr_3$ adopt the perovskite structure, that both are distorted at room temperature — tetragonally and monoclinically, respectively, in his assignment — and that each transforms to the ideal cubic aristotype on heating, at 47 and 130 °C. Møller also reported photoconductivity, tunable across the visible by choice of halide. Two elements of the modern picture are therefore present from the outset: a cubic reference structure that is *not* the ground state but a high-temperature average, and a semiconducting gap in the visible controlled by the halide. Both observations had to be rediscovered.

The hybrid organic-inorganic compounds followed two decades later. Weber[23] synthesized $CH_3NH_3PbX_3$ for X = Cl, Br, I, refined cubic lattice constants of 5.68, 5.92, and 6.27 Å respectively, and noted the intense colouration of the bromide and iodide — black, in the case of the iodide, a fact whose significance for light harvesting would wait thirty years. He rationalized the bonding in terms of what he termed *p-resonance bonding*: a delocalized description of the Pb-X network in which the frontier states are shared across the framework. Read today, this is a qualitative anticipation of the antibonding, delocalized band-edge character that Section 5 develops in modern language, and it is a small historical irony that the earliest paper on the hybrid compounds contains the germ of the argument now invoked to explain their defect behaviour. Weber's compounds attracted no immediate attention as electronic materials.

## 2.2. The dormant era, 1985-2003: dielectrics, dynamics, and excitons

The two decades that followed produced, almost entirely outside the semiconductor community, a body of work that anticipates much of the present discussion. Three strands are worth separating, because each terminates in a modern controversy.

### 2.2.1. A-site cation dynamics

Wasylishen, Knop, and Macdonald[26] established by nuclear magnetic resonance that the methylammonium cation is not statically ordered but reorients rapidly within its cuboctahedral cavity. Poglitsch and Weber[27] followed with millimetre-wave dielectric spectroscopy, resolving the orientational contribution to the dielectric function directly and characterizing its picosecond-scale relaxation; Onoda-Yamamuro, Matsuo, and Suga[28] mapped the associated phase-transition sequence calorimetrically, identifying the entropy released at each transition with the successive unlocking of orientational degrees of freedom. The structural counterpart was supplied much later, when neutron powder diffraction across the full temperature range located the transitions of $MAPbI_3$ at 165 K (orthorhombic to tetragonal) and 327 K (tetragonal to cubic) and imaged the cation directly: ordered in the orthorhombic ground state, disordered over four sites in the tetragonal phase, and effectively free in the cubic phase[29,30]. By 1990, however, the essential point was already established: $MAPbX_3$ is an orientationally disordered crystal in its high-temperature phases, in the same sense as the plastic crystals of the molecular-solids literature. The relevance of this to charge transport was not pursued, for the excellent reason that no one was measuring charge transport.

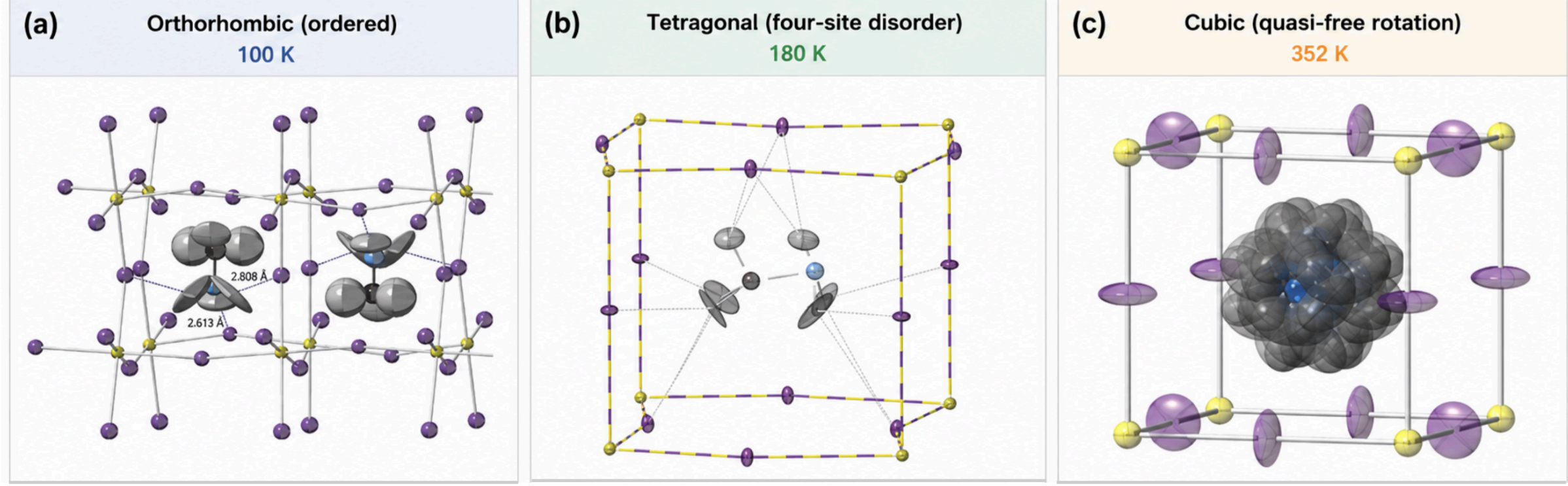

**Fig. 6 | Neutron powder diffraction evidence for the orientational state of the methylammonium cation in $MAPbI_3$**: hydrogen atomic displacement parameter (ADP) ellipsoids compared across the (a) orthorhombic (ordered), (b) tetragonal (four-site disorder), and (c) cubic (quasi-free rotation) phases[29]. Reproduced from Weller et al.[29] under the terms of the Creative Commons Attribution 3.0 Unported Licence.

This strand deserves emphasis because a significant fraction of the post-2013 literature attributed the anomalous behaviour of the material to the molecular dipole of the MA cation — rotating dipoles were invoked to screen carriers, to form ferroelectric domains, and to assist charge separation. Yaffe *et al.*[3] subsequently demonstrated that the defining low-frequency Raman signature — the strong quasielastic central peak of the cubic phase — appears equally in the all-inorganic $CsPbBr_3$, whose A-site cation is a monatomic ion with no dipole and no orientational degree of freedom, and is therefore a property of the lead-halide framework rather than of the organic cation. The molecular dynamics established in the 1980s is real, and it contributes to the dielectric response at low frequency (Section 6); but it is not the origin of the anomaly. We adopt this as a fixed point of interpretation throughout the review.

### 2.2.2. Excitons and magneto-optics

Hirasawa *et al.*[22] measured the magnetoabsorption of the lowest exciton in $MAPbI_3$ in pulsed fields, and Tanaka *et al.*[31] provided a comparative study of the excitons in $MAPbBr_3$ and $MAPbI_3$. These experiments established the essential facts — a hydrogenic Wannier-Mott exciton with a binding energy of a few tens of meV in the low-temperature phase — a full decade before the photovoltaic literature reopened the question. The reopened question then went badly: between 2013 and 2015, reported binding energies for $MAPbI_3$ spanned roughly 2 to 60 meV, depending on whether they were extracted from absorption lineshape fits, temperature-dependent photoluminescence, or dielectric arguments, and on which value of the dielectric constant — high-frequency, static, or something between — the analysis assumed. The controversy was ultimately settled not by new physics but by better magnets: pulsed fields to 150 T permitted resolution of the excited exciton states and of the free-carrier Landau fan, and hence a determination of the reduced mass, $\mu$ = 0.104 plus or minus 0.003 free-electron masses, that does not depend on any assumed dielectric constant, yielding a binding energy of approximately 16 meV in the low-temperature phase and a few meV smaller at room temperature[1,32]. The episode is instructive twice over: methodologically, as a case in which a model-independent measurement ended a dispute that model-dependent measurements had sustained; and physically, because the *reason* the intermediate values were so scattered — the frequency dependence of the screening — is itself the physics of Section 6.

"Figure adapted from [2], reproduced with permission in the final published version"

**Fig. 7 | Magneto-absorption fan diagram of $MAPbI_3$ in pulsed fields to 150 T**: exciton and free-carrier Landau levels splitting with field, from which the model-independent reduced mass and binding energy of Section 2.2.2 and Section 7.1 are extracted. Reproduced with permission from Miyata et al.[1], © 2015 Springer Nature.

### 2.2.3. Layered systems

The two-dimensional derivatives have their own, largely independent lineage. Mitzi *et al.*[33] reported conducting layered organic-inorganic halides built from perovskite sheets, demonstrating that the electronic framework survives dimensional slicing and that the organic gallery is a tunable, electronically passive spacer. Through the 1990s a substantial literature developed on these natural multiple-quantum-well structures — exciton binding energies enhanced to hundreds of meV by the combination of quantum confinement and the dielectric mismatch between well and barrier, the image-charge ("dielectric confinement") effect that has no analogue in epitaxial III-V wells[18]. Kondo *et al.*[34] examined the localization of electronic states in amorphized $CsPbBr_3$ and $CsPbCl_3$, an early probe of what disorder does to this band structure. It is a notable asymmetry of the field that the layered compounds were understood as semiconductor heterostructures before the three-dimensional parents were understood as semiconductors at all; Section 10 inherits this literature directly.

## 2.3. Photovoltaic ignition, 2009-2013

Kojima *et al.*[35] employed $MAPbX_3$ nanoparticles as visible-light sensitizers in a liquid-electrolyte photoelectrochemical cell. The devices were unstable — the electrolyte dissolved the perovskite — the efficiency modest, and the report attracted limited attention for three years. The transition to all-solid-state architectures changed this decisively: Kim *et al.*[36] reported a submicron mesoscopic cell exceeding nine per cent, and Lee *et al.*[37], replacing the mesoporous titania electron acceptor with an inert alumina scaffold, exceeded ten per cent while demonstrating in passing that the perovskite transports electrons perfectly well by itself — the observation that first hinted the material was not a dye but a semiconductor. Within three years the certified efficiency had passed twenty per cent, a rate of improvement without precedent in photovoltaics. We compress the device history deliberately: its details belong to a different review, and its role here is only to explain the sudden mobilization of the community whose physics output the remaining sections analyses.

For the purposes of this review the decisive publication of this period is not a device paper but a transport measurement. Stranks *et al.*[6] determined ambipolar electron-hole diffusion lengths exceeding one micrometre in solution-processed films of the mixed-halide compound. This is the observation that converted a device result into a physics problem, because a diffusion length of that magnitude in a material of that structural quality is not accounted for by any conventional semiconductor argument: with the mobilities then being measured, it implies carrier lifetimes approaching the microsecond scale in the presence of defect densities that should, on Shockley-Read-Hall grounds, extinguish the carriers in nanoseconds. Everything discussed in Sections 6 through 9 is, in one way or another, an attempt to explain this number.

## 2.4. The return to fundamental physics, 2013-present

The subsequent decade may be divided into three overlapping phases, which we present in the order of their opening rather than as a strict sequence.

### 2.4.1. Relativistic electronic structure, from 2013

Even *et al.*[38] established that spin-orbit coupling is not a refinement but a leading-order effect in these compounds: on the Pb 6p conduction states it produces a splitting of roughly an electronvolt, comparable to the gap itself, and non-relativistic band structures are meaningless despite their fortuitous numerical agreement with measured gaps — an agreement produced by the near-cancellation of two large errors, the neglected spin-orbit stabilization of the conduction band and the usual density-functional underestimate of the gap. The cancellation misled a number of early studies, and its exposure is a standing caution recorded in Appendix A. Brivio *et al.*[39] supplied the relativistic quasiparticle self-consistent GW treatment that removes both errors at once and identified Rashba-Dresselhaus spin splittings at the band extrema in symmetry-broken configurations — the entry point of the spin physics of Section 11.

### 2.4.2. Electron-phonon coupling and polarons, from 2015

Zhu and Podzorov[10] proposed that carriers are protected as Fröhlich large polarons, a hypothesis framed explicitly as an explanation of the coexistence of modest mobility with long lifetime. Wright *et al.*[13] quantified the coupling spectroscopically, identifying the dominant longitudinal optical modes near 11.5 and 15.3 meV and demonstrating from the temperature dependence of the emission linewidth that Fröhlich scattering dominates at room temperature, with acoustic-phonon scattering negligible — the quantitative basis for the intrinsic-mobility analysis of Herz[2] and Section 8. Zhu *et al.*[11] reported time-resolved evidence that energetic carriers are screened on the timescale of the orientational and lattice relaxation, and Miyata *et al.*[12] measured polaron formation times of order a picosecond in both $MAPbBr_3$ and $CsPbBr_3$ — again the hybrid/inorganic comparison that recurs as the field's principal control experiment. The many-body machinery was placed on a first-principles footing by Schlipf, Poncé, and Giustino[40].

### 2.4.3. Anharmonicity and dynamic disorder, from 2017

Yaffe *et al.*[3] identified the anharmonic central peak as intrinsic to the lead-halide framework, as discussed in Section 2.2.1. Ferreira *et al.*[41] demonstrated the weak dispersion and strong damping of the optical branches directly by inelastic scattering. Lanigan-Atkins *et al.*[4] resolved the momentum structure of the overdamped dynamics in $CsPbBr_3$: the low-energy fluctuations are correlated head-to-tail octahedral rotations confined to two-dimensional sheets, appearing as rods of diffuse intensity along the cubic zone edges, a structure the authors identify as a possible precursor to two-dimensionally confined polarons. On the theoretical side, Mayers *et al.*[19], Schilcher *et al.*[20], and Schilcher *et al.*[42] developed the description of carrier transport in a lattice whose fluctuations are large, correlated, and slow, while Zhao *et al.*[21] argued that the cubic phase must be treated as *polymorphous* — a thermal distribution of locally symmetry-broken configurations sharing a cubic average — rather than as a genuinely cubic crystal, with quantitative consequences for gaps, effective masses, and spin splittings. The thermal-transport counterpart of the same softness had been established earlier: room-temperature lattice thermal conductivities near 0.5 W $m^{-1}$ $K^{-1}$ in $MAPbI_3$ (ref. 43) and comparably low values in the all-inorganic compounds[44], among the lowest known for fully dense crystalline solids and the quantitative content of the phonon-glass half of the crystal-liquid duality.

Running alongside these developments is a critical counter-current that this review treats as an equal partner rather than an afterthought. Egger *et al.*[15] subjected the accumulated claims of anomaly to systematic scrutiny and found many wanting. Zhang *et al.*[8] argued from first-principles capture calculations that halide perovskites suffer defect-assisted nonradiative recombination at rates comparable to conventional semiconductors, so that defect tolerance in the strong sense is not supported. Mosquera-Lois *et al.*[9] surveyed the competing models of defect tolerance and found each to be partial. Section 9 treats this dispute as unresolved, which it is.

## 2.5. What the history establishes

Three lessons follow from this sequence and inform the structure of what follows.

First, the anomalies of this material are anomalies of the inorganic framework, not of the organic cation. The chronology misled the field on this point for several years, because the hybrid compounds were studied first and the all-inorganic controls came later; each time the control experiment was performed — central peak, polaron formation time, ultralow thermal conductivity — the caesium compound reproduced the behaviour of the hybrid. Where a property survives the substitution of Cs for MA, the molecular degree of freedom is not its cause. The converse also holds and is easily forgotten properties that do *not* survive the substitution, principally in the low-frequency dielectric response, are genuinely molecular in origin, and Section 6 keeps the two contributions separated.

Second, several questions treated as new after 2013 had been posed, and partially answered, in the dielectric and excitonic literature of the preceding decades. The exciton binding-energy controversy is the clearest case: it was resolved by returning to the technique of Hirasawa *et al.*[22] with a stronger magnet. A field moving quickly is entitled to rediscover things; it is not entitled to leave them uncited, and part of the purpose of Section 2.2 is restitution.

Third, and most consequentially for the organization of this review, the historical progression itself moved from the electronic to the structural. The field began with band structures and defect levels, encountered discrepancies — wrong sign of the gap's temperature coefficient, scattered binding energies, mobilities well below the predictions of harmonic scattering theory — and was driven, measurement by measurement, towards lattice dynamics. Having arrived at that conclusion, we see no reason to make the reader repeat the journey, and we begin instead where the field ended up: with the structure and dynamics of the lattice, in Sections 3 and 4.

# 3. Crystal Structure and Symmetry

This section establishes the structural vocabulary on which everything later depends: the cubic aristotype and its bonding; the geometric rules that decide which compositions adopt it; the classification of the octahedral tilts that lower its symmetry; the phase sequences of the principal compounds; and — the point at which this section hands over to Section 4 — the growing body of evidence that the high-temperature "cubic" phase is cubic only on average, so that the crystallography of this material cannot be separated from its dynamics.

## 3.1. The aristotype and its bonding

The ideal perovskite structure, space group Pm-3m with one formula unit per cell, is generated by a single geometric idea: corner-sharing octahedra. The $Pb^{2+}$ cation sits at the centre of a $PbX_6$ octahedron; each halide is shared between exactly two octahedra, so that the framework is a three-dimensional network of Pb-X-Pb bridges with an ideal bridge angle of 180 degrees; and the A-site cation occupies the cuboctahedral cavity between eight octahedra, twelve-fold coordinated by halides. Two consequences of this topology run through the entire review. First, the framework is *open*: the cavity is large, the coordination of the A cation is loose, and low-energy rigid-unit motions — rotations of essentially undistorted octahedra about their shared corners — exist that cost little elastic energy. These rigid-unit modes are the structural raw material of every phase transition in Section 3.4 and of the anharmonic dynamics of Section 4. Second, the electronically active states live on the framework, not on the A cation. The valence-band maximum is the antibonding Pb 6s - X np combination and the conduction-band minimum is dominantly Pb 6p (Section 5); the A-site cation contributes no states near the gap and acts, electronically, as a charge-compensating spacer. This division of labour — framework for electrons, cavity for structural chemistry — is what makes the A-site substitution MA/FA/Cs such a clean experimental control, and it is also why the band gaps of the hybrid and all-inorganic compounds are so similar despite entirely different cavity chemistry.

The bond itself deserves a comment, because its character sets the energy scales. The Pb-X interaction is substantially ionic but with a covalent component that increases from chloride to iodide; the bond is long, soft, and highly polarizable, and the $Pb^{2+}$ ion carries the $6s^2$ lone pair whose stereochemical activity — the tendency of the s-p hybridized lone-pair density to push the cation off-centre — supplies an electronic driving force for local symmetry breaking that coexists with, and in the cubic phase competes against, the steric preferences of the framework. The lone pair appears three times in this review in three different roles: here, as a structural actor; in Section 5, as the origin of the antibonding valence-band character; and in Section 4, as a contributor to the anharmonic potential landscape. The unity of these three roles is part of the organizing hypothesis of Section 1.

## 3.2. Which compositions form the structure: tolerance and octahedral factors

The classical answer to the question "which A, B, X combine into a perovskite" is geometric. Goldschmidt[45] observed that the ideal structure requires the A-X and B-X bond lengths to be commensurate — the A cation must fill the cavity that the framework provides — and encoded the condition in the tolerance factor $t$, built from the ionic radii of the three sites:

$$t = (r_A + r_X) / [\sqrt{2}\,(r_B + r_X)] \tag{2}$$

A perfect fit corresponds to $t = 1$; empirically, perovskites form for roughly $0.8 < t < 1.0$, with the lower part of the range producing strongly tilted, lower-symmetry variants; below about 0.8 the A cation is too small to prop the cavity open and lower-connectivity structures win, while above 1.0 the cation is too large for the cubic cage and face-sharing hexagonal stacking or layered motifs are favoured[46]. A second condition, the octahedral factor μ, expresses the requirement that the B cation fit its own octahedron:

$$\mu = r_B / r_X \tag{3}$$

with μ greater than approximately 0.41 required for octahedral coordination. Figure 8 summarizes the map.

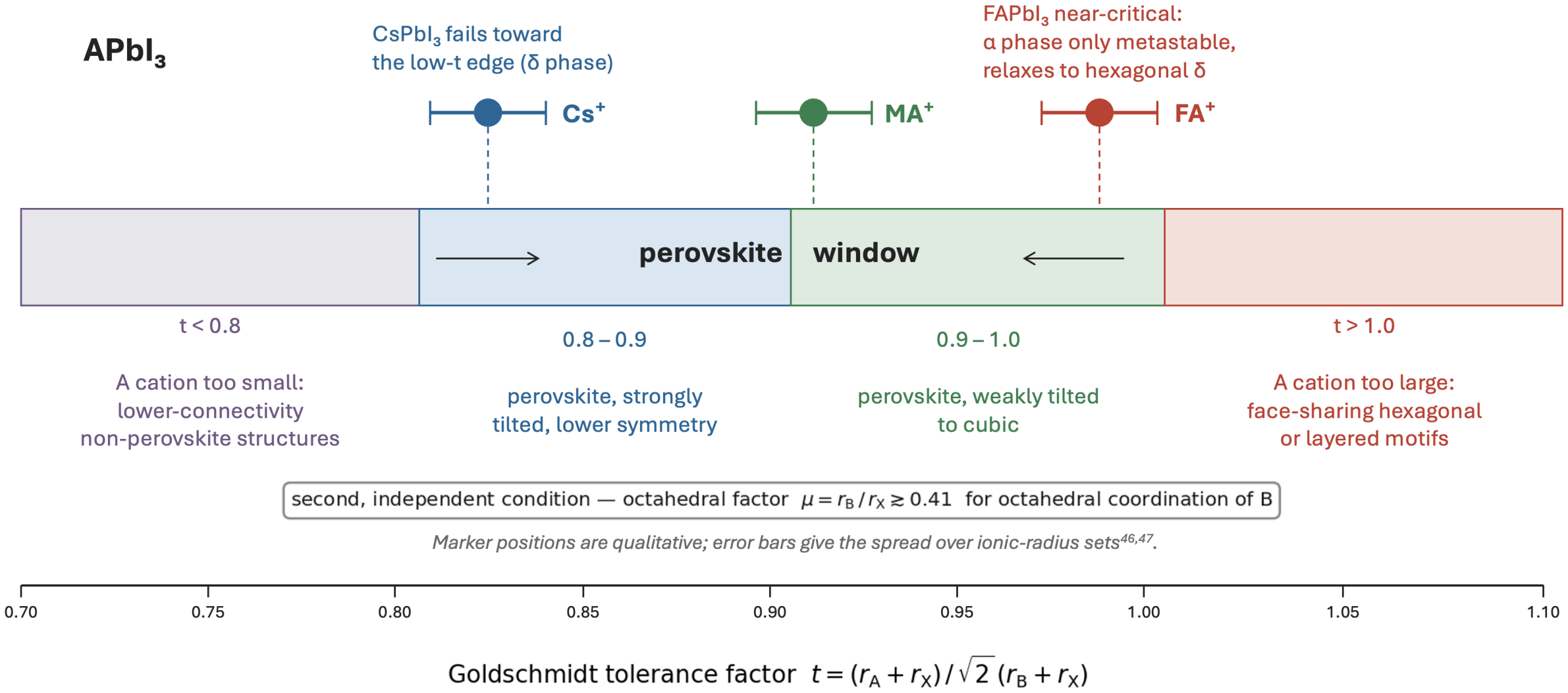

**Fig. 8 | Schematic structure map in the Goldschmidt tolerance factor.** Shaded regions indicate the structural outcomes empirically associated with each range of t; markers place the three principal A cations of this review for the iodide framework,

with error bars indicating the spread produced by different choices of ionic-radius set. The placements — Cs below MA below FA, with FA close to unity — are robust; the absolute values are not, and quantitative claims should not be hung on them[46,47].

Extending the tolerance factor to molecular cations requires assigning them a radius, which is a modelling decision rather than a measurement. Kieslich, Sun, and Cheetham[48,49] proposed effective radii for freely rotating molecular ions — treating the rotationally averaged cation as a sphere — and used them to survey thousands of hypothetical hybrid compositions; on this scale MA sits comfortably inside the perovskite window of the iodide framework while FA sits near its upper edge. The near-criticality of FA is physical, not an artefact of the bookkeeping: $FAPbI_3$ at room temperature is only metastable in its cubic perovskite α phase, whose average structure was established by powder neutron diffraction [29], and relaxes over hours to days into a non-perovskite hexagonal δ phase in which the octahedra share faces — precisely the failure mode the map predicts for t at or above unity. $CsPbI_3$ sits near the opposite edge and fails in the opposite direction, its perovskite phases metastable at room temperature against a lower-connectivity δ phase. That the two technologically central iodides bracket the stability window from its two sides is a structural fact with device consequences we do not pursue, but it is also a warning that the perovskite phase of this family is, over much of composition space, a fragile object stabilized as much by entropy — including the rotational entropy of the molecular cation — as by energy.

The limits of the geometric description should be stated with equal clarity. Applied naively across the halide families, the Goldschmidt factor misclassifies a substantial fraction of compounds — its predictive accuracy degrades systematically from chlorides to iodides as the bonding becomes more covalent and the hard-sphere picture less appropriate — and improved descriptors constructed by data-driven methods, such as the tau factor of Bartel *et al.*[47], outperform it precisely by weakening the hard-sphere assumption. For the purposes of this review the tolerance factor is an organizing heuristic and a language for trends, not a law; where we invoke it, nothing quantitative depends on it.

## 3.3. Octahedral tilts and the Glazer classification

Between the ideal cubic structure and wholesale reconstruction lies the family of distortions that actually occur cooperative rotations of essentially rigid octahedra about the cubic axes. Because the octahedra share corners, a rotation of one about a given axis forces its four in-plane neighbours to counter-rotate; the only freedom left is the *stacking phase* — whether successive layers along the rotation axis rotate in the same sense or in alternation. Glazer[50] turned this observation into a notation that the field has used ever since: a tilt system is written as three symbols, one per cubic axis, each carrying a superscript 0, +, or - according to whether the rotation about that axis is absent, in-phase from layer to layer, or out-of-phase, with repeated letters denoting equal tilt magnitudes. Thus, $a^0a^0a^0$ is the untilted aristotype; $a^0a^0c^+$ is a single in-phase tilt about one axis; $a^0a^0c^-$ the corresponding out-of-phase tilt; $a^-b^+a^-$ the three-tilt system of the orthorhombic ground states below. Glazer enumerated twenty-three such systems; subsequent group-theoretical analysis reduced the crystallographically distinct ones to fifteen and supplied the machinery — irreducible representations of the Pm-3m parent — that connects each tilt system to the space group it generates[51]. Figure 9 shows the three patterns that matter most in this family.

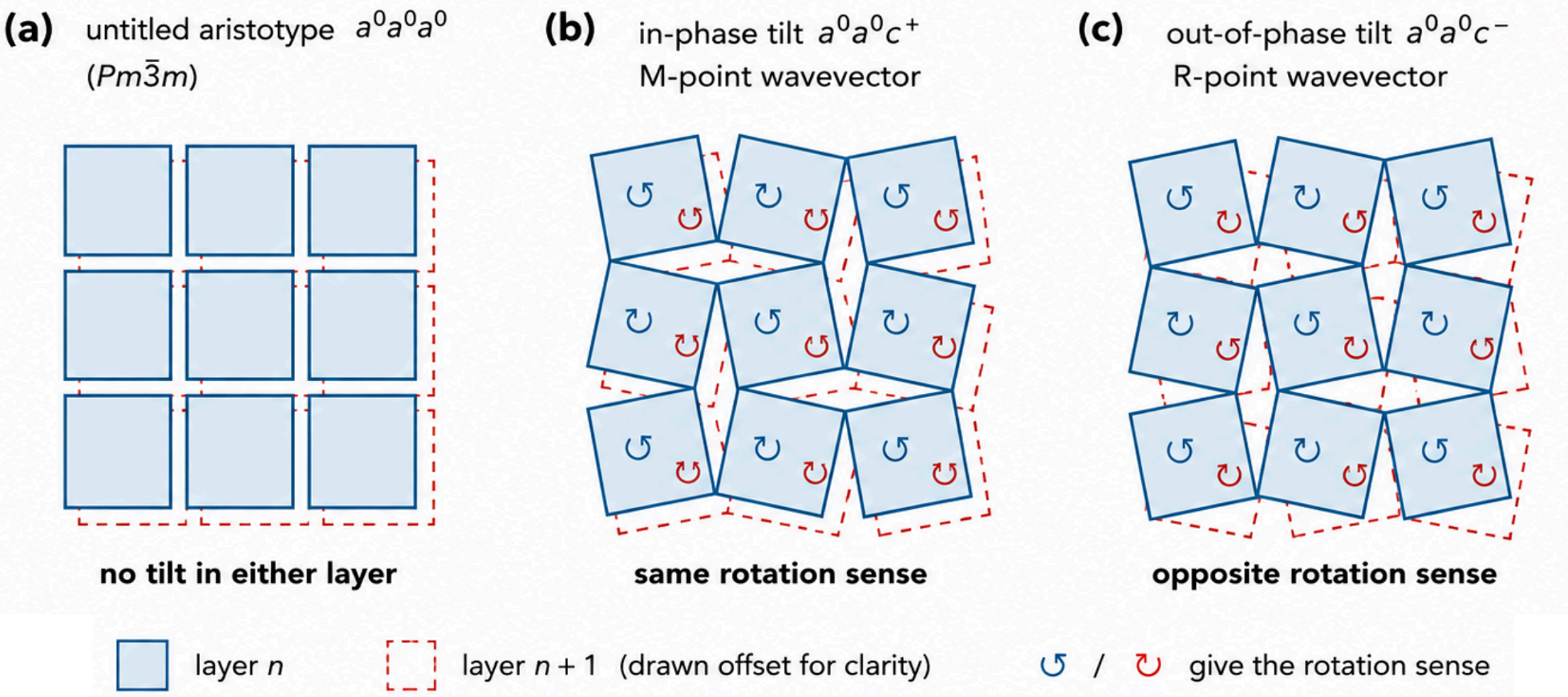

**Fig. 9 | The elementary tilt patterns, viewed along the tilt axis; each square is the projection of one $PbX_6$ octahedron, solid and dashed marking successive layers along the viewing direction.** (a) Untilted aristotype. (b) In-phase tilt $a^0a^0c^+$: successive layers rotate in the same sense; the pattern is periodic with the wavevector of the cubic M point. (c) Out-of-phase tilt $a^0a^0c^-$: successive layers alternate; the pattern carries the wavevector of the R point. In-plane counter-rotation of neighbouring octahedra, common to (b) and (c), is enforced by corner sharing.

The connection to lattice dynamics is direct and is the bridge on which Section 4 is built. In the phonon spectrum of the cubic phase, the in-phase tilt is the eigenvector of a zone-boundary mode at the M point (irreducible representation $M_3^+$) and the out-of-phase tilt of a mode at the R point ($R_4^+$); a phase transition into a tilted structure is, in soft-mode language, the condensation of the corresponding zone-boundary phonon, whose frequency falls toward zero as the transition is approached from above. For $CsPbCl_3$, the cubic-to-tetragonal transition is precisely the condensation of the $M_3^+$ mode, carrying Pm-3m into P4/mbm with tilt system $a^0a^0c^+$. In the lead halides, however — and this is the fact that separates them from the textbook soft-mode ferroelastics and that Section 4 develops at length — the modes in question are not underdamped oscillators that soften gracefully; over wide temperature ranges above the transitions they are overdamped, and entire branches along the M-R zone edge lie low and flat, so that the "soft mode" is better described as a continuum of slow, large-amplitude, strongly anharmonic rotational fluctuations [4]. The Glazer classification survives this complication — it classifies the *instantaneous* local patterns as well as the condensed ones — but the mean-field soft-mode narrative does not, and we flag that here so that the phase diagram of the next subsection is read with the right reservations.

What drives the tilting is a question with a layered answer. The zeroth-order driver is steric, as the tolerance factor already encodes: an undersized A cation leaves the halides underbonded, and rotating the octahedra sweeps the halides inward to improve the A-X coordination, which is why the tilt amplitudes grow as t decreases and why $CsPbX_3$ tilts more strongly than $MAPbX_3$. At the next order the driver is electronic: first-principles analysis finds that the tilts additionally gain energy from improved orbital hybridization — a second-order Jahn-Teller mechanism involving the lone pair — and, in the hybrid compounds, from hydrogen bonding between the ammonium head of the cation and the halides, without which, in the case of $MAPbI_3$, the calculated tilting propensity largely disappears[52]. The hydrogen-bonding contribution couples the cation orientation to the tilt pattern and is the microscopic root of the order-disorder character of the hybrid transitions discussed next.

## 3.4. Phase sequences of the principal compounds

The three-dimensional perovskites considered here generally share a qualitative sequence on cooling — a cubic average structure followed by a tetragonal phase with a single dominant tilt and, for the compounds that stabilize the conventional perovskite Pnma ground state, an orthorhombic phase with the three-tilt system $a^-b^+a^-$ (space group Pnma) — but the details differ in ways that carry physical information. $FAPbI_3$ is an important exception: its thermodynamic ground state is the non-perovskite hexagonal δ phase, whereas the perovskite α phase is metastable at room temperature. Figure 10 therefore focuses on the two best-characterized systems, $MAPbI_3$ and $CsPbBr_3$.

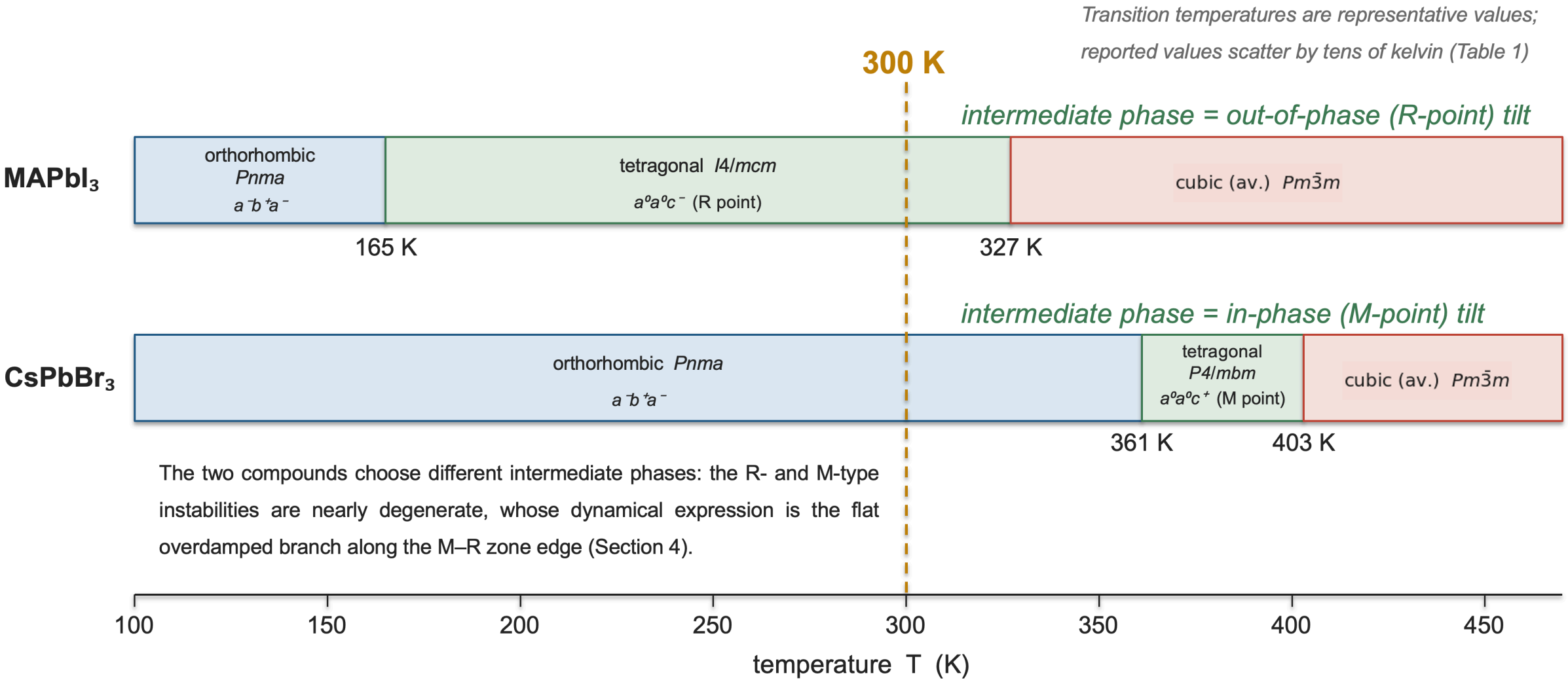

**Fig. 10 | Phase sequences of $MAPbI_3$ (transitions[29,30] at 165 and 327 K) and $CsPbBr_3$ (361 and 403 K; see text), with space groups and Glazer tilt systems.** The two compounds choose *different* intermediate phases: the tetragonal phase of $MAPbI_3$ is the out-of-phase (R-point) tilt, that of $CsPbBr_3$ the in-phase (M-point) tilt — evidence that the R- and M-type instabilities are nearly degenerate in this family, a near-degeneracy whose dynamical expression is the flat overdamped branch along the M-R zone edge (Section 4). "cubic (av.)" emphasizes that the high-temperature phase is cubic as a time and space average only (Section 3.5).

$MAPbI_3$ passes from the orthorhombic ground state to the tetragonal phase (I4/mcm, $a^0a^0c^-$) at 165 K and to the cubic average phase at 327 K; the assignments rest on single-crystal and powder diffraction[53] and on the full-range neutron studies of Weller *et al.*[29] and Whitfield *et al.*[30], the latter characterizing the tetragonal-cubic transition as close to tricritical — at the border between continuous and first-order character. The transitions are not merely displacive: they are simultaneously ordering transitions of the molecular cation, which is orientationally frozen in the orthorhombic phase, hops among four sites in the tetragonal phase, and is quasi-free in the cubic phase (Section 2.2.1), with the calorimetric entropies of Onoda-Yamamuro, Matsuo, and Suga[28] quantifying the successive unlocking. The coupling term is the hydrogen bond of Section 3.3: cation order and tilt order are not independent order parameters but linked ones, and the hybrid transitions are best described as coupled order-disorder/displacive events — a hybrid character that the all-inorganic compounds, lacking the orientational degree of freedom, do not share.

$CsPbBr_3$ exhibits the same endpoint phases with different topology in between: orthorhombic Pnma up to 361 K, then the *in-phase* tetragonal phase P4/mbm ($a^0a^0c^+$), then the cubic average phase above 403 K. The comparison with $MAPbI_3$ is more instructive than either sequence alone. That two members of the same family, with the same framework and the same ground state, select opposite intermediate tilts says that the R-point and M-point instabilities are nearly degenerate — the free-energy landscape in the space of tilt patterns is shallow and nearly flat along the line connecting them — and this near-degeneracy is exactly what the inelastic-scattering experiments see dynamically as a low, flat, overdamped branch spanning the entire M-R edge of the cubic zone (Lanigan-Atkins *et al.*, 2021; Section 4). $CsPbCl_3$, historically the first case, condenses the $M_3^+$ mode at its cubic-tetragonal transition near 320 K, the transition itself established by Møller[24] and the mode assignment by the neutron-scattering study of Fujii, Hoshino, Yamada, and Shirane[93]. $FAPbI_3$ adds the complication of Section 3.2: its perovskite α phase, cubic on average at room temperature[29], competes with the non-perovskite hexagonal δ phase, so that its low-temperature behaviour is governed as much by phase competition as by tilt condensation, and we return to it only where that competition illuminates the physics.

**Table 1.** Phase sequences of the principal three-dimensional compounds discussed in Section 3.4. Transition temperatures for the hybrid and the all-inorganic bromide scatter by up to tens of kelvin between reports (Section 3.4); the values below are representative, not universal.

| Compound | Ground state | Intermediate phase | Cubic (average) onset | Refs |
|---|---|---|---|---|
| $MAPbI_3$ | Orthorhombic *Pnma* ($a^-b^+a^-$) | Tetragonal *I*4/*mcm* ($a^0a^0c^-$), 165–327 K | *Pm*-3*m* above 327 K | 29,30 |
| $CsPbBr_3$ | Orthorhombic *Pnma* ($a^-b^+a^-$), below 361 K | Tetragonal *P*4/*mbm* ($a^0a^0c^+$), 361–403 K | *Pm*-3*m* above 403 K | Sec. 3.4 |
| $CsPbCl_3$ | Tetragonal at room temperature (Møller 1958 assignment) | — | *Pm*-3*m* above ≈320 K ($M_3^+$ mode condensation) | 24,93 |
| $FAPbI_3$ | Non-perovskite hexagonal δ phase (thermodynamic ground state) | — | Perovskite α phase metastable at room temperature; cubic *Pm*-3*m* on average | 29 |

One quantitative habit is worth flagging as a caution for the reader who consults the primary literature. Transition temperatures in this family scatter by up to tens of kelvin between reports — sample stoichiometry, strain state, crystallite size, and heating versus cooling all shift them — and the $CsPbBr_3$ values quoted above (361 and 403 K) are representative rather than universal. Nothing in this review depends on their precise values; what matters is the sequence, the tilt assignments, and the near-degeneracy argument.

## 3.5. Average versus instantaneous structure

The final structural fact is the one that dissolves the boundary between this section and the next. Bragg diffraction measures the time- and space-averaged density, and by that measure the high-temperature phase is cubic. But three independent lines of evidence show that the *instantaneous* local structure is not. First, the refined atomic displacement parameters of the halides in the cubic phase are large and grossly anisotropic — pancaked perpendicular to the Pb-X bond — exactly as expected if the octahedra are executing large-amplitude rotational excursions about an untilted mean[29,30]. Second, diffuse scattering, which probes correlations rather than averages, reveals structured intensity — the two-dimensionally correlated rods along the zone edges — demonstrating that the local tilts are not independent site noise but organized, slowly fluctuating patterns [4]. Third, first-principles thermodynamics reaches the same conclusion from the other side: allowing a nominally cubic supercell to explore its energy landscape, Zhao *et al.*[21] find that the ideal cubic geometry is not a minimum but a saddle-like average over a thermal distribution of locally tilted, lower-symmetry configurations — a situation they name *polymorphous* — and that computing band gaps, effective masses, and spin splittings on the ideal average geometry rather than over the distribution introduces errors that are not small (Section 5).

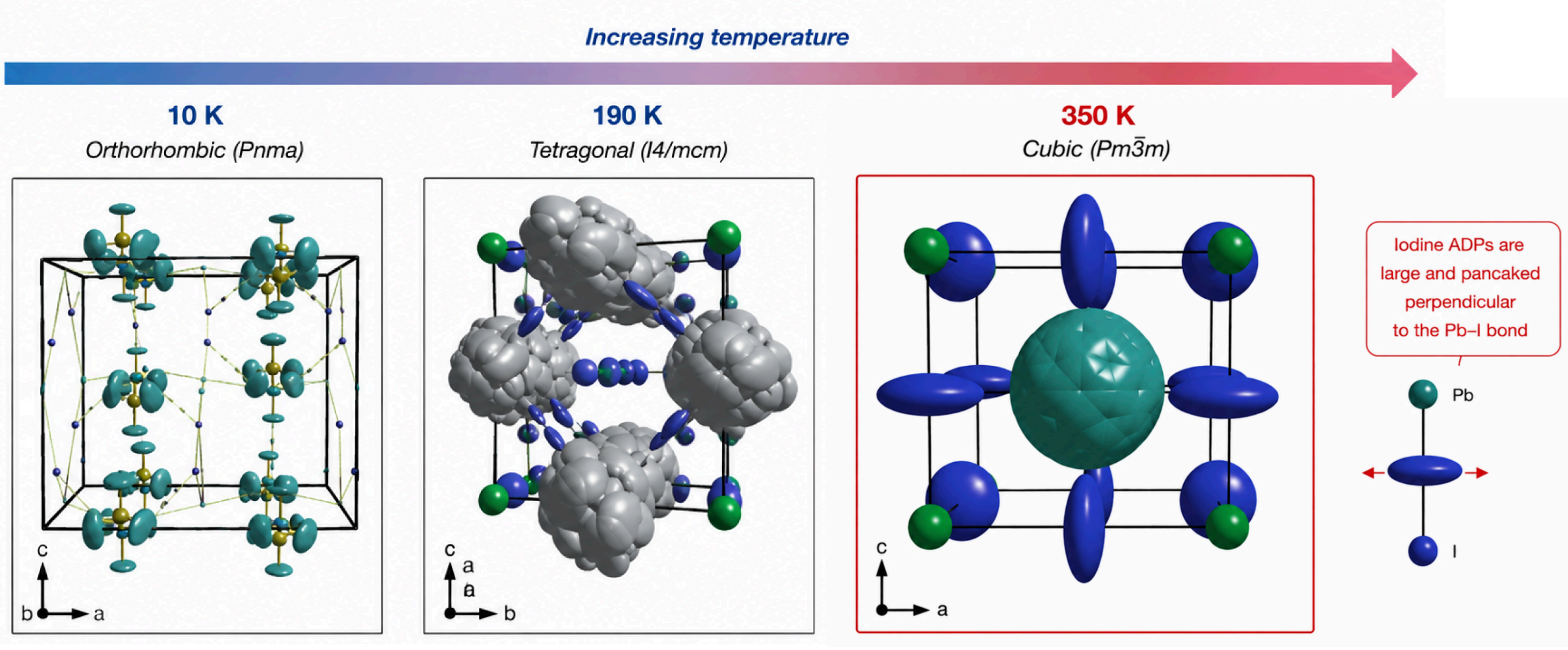

**Fig. 11 | Anisotropic atomic displacement parameters of the halides in cubic $MAPbI_3$, from full-temperature-range neutron powder diffraction:** the halide ADPs are large and pancaked perpendicular to the Pb-X bond, the structural signature of large-amplitude octahedral rotational excursions about an untilted mean. Reproduced from Whitfield et al.[30] under the terms of the Creative Commons Attribution 4.0 International License (CC BY 4.0).

“Figure adapted from [19, 21], reproduced with permission in the final published version”

**Fig. 12 | Polymorphous and dynamic origins of electronic-structure fluctuations in nominally cubic halide perovskites.** (a) Locally distorted, lower-symmetry configurations in a polymorphous cubic network, showing distributions of octahedral tilting, B-site displacement and octahedral volume.[21] (b) Electronic-structure differences between monomorphous and polymorphous descriptions, including relative band-gap ordering and the joint density of states (JDOS).[21] (c) Instantaneous band-gap fluctuations along a molecular-dynamics trajectory[19]. Panels (a) and (b) reproduced with permission from Zhao et al., Phys. Rev. B 101, 155137 (2020). © 2020 American Physical Society. Panel (c) reproduced with permission from Mayers et al., Nano Lett. 18, 8041-8046 (2018). © 2018 American Chemical Society.

The conceptual consequence deserves to be stated as the section's conclusion, because it licenses the ordering of the whole review. For this family, a phase label such as "cubic Pm-3m" is a statement about a *distribution*, not about a geometry: the symmetry group describes the statistics of the fluctuations, not any configuration the crystal visits. Static crystallography, applied at face value, therefore delivers a fictitious reference structure — and any theory built on that reference, from band structures to defect levels to Rashba physics, inherits the fiction unless the fluctuations are put back in. How large the fluctuations are, how fast, how correlated, and what it costs to ignore them: those are dynamical questions, and they are the subject of Section 4.

# 4. Lattice Dynamics and Anharmonicity

This is the load-bearing section of the review's first half. Section 3 ended with the claim that the crystallography of this family cannot be separated from its dynamics; here we make that claim quantitative. The section follows the standard four-part logic: the experimental facts (4.1-4.2), the harmonic framework and its partial successes (4.3), the specific ways in which that framework fails (4.4), and the consequences and open questions (4.5-4.6).

## 4.1. The experimental facts: overdamping, central peaks, and correlated rotations

Three experiments define the problem. First, low-frequency Raman scattering. In the orthorhombic ground state, the spectra of $MAPbX_3$ and $CsPbBr_3$ show the sharp folded-phonon lines of an ordinary crystal; on heating through the tetragonal and into the cubic phase, the sharp lines broaden, merge, and are progressively replaced by a strong quasielliptical central peak whose wings extend over the entire low-frequency region[3]. The central peak is present, and comparably strong, in the hybrid and in the all-inorganic compound; whatever produces it therefore lives in the $[PbX_3]^-$ framework. Its intensity and width grow with temperature in a manner characteristic of relaxational — not oscillatory — dynamics, and Yaffe *et al.* interpret it as the signature of large-amplitude local polar fluctuations of the lead-halide cage.

Second, momentum-resolved inelastic and diffuse scattering. In $CsPbBr_3$, Lanigan-Atkins *et al.*[4] find that the low-energy fluctuation spectrum of the cubic phase is concentrated along the M-R edges of the Brillouin zone as rods of diffuse intensity, with the corresponding modes overdamped — linewidths exceeding their frequencies — over a wide temperature window. The real-space content of a zone-edge rod is specific: the octahedral rotations are strongly correlated *within* two-dimensional planes (head-to-tail, as the corner-sharing constraint of Section 3.3 demands) and essentially uncorrelated *between* planes. The cubic phase of this material is thus host to a liquid-like, quasi-two-dimensional rotational texture, dynamically disordered along the stacking direction. The same experiments show conventional, dispersive acoustic branches at long wavelength: the crystal is elastically a solid and rotationally a liquid at the same time.

Third, the optical branches themselves. Ferreira *et al.*[41] demonstrate by inelastic scattering on hybrid crystals that the optical phonons are weakly dispersive and strongly damped even where they remain underdamped — flat bands of slow, spatially localizable vibrations rather than propagating waves. Together with the ultralow lattice thermal conductivities discussed below, this completes a picture in which the harmonic quasiparticle — the phonon — survives cleanly only in the acoustic sector and at low temperature.

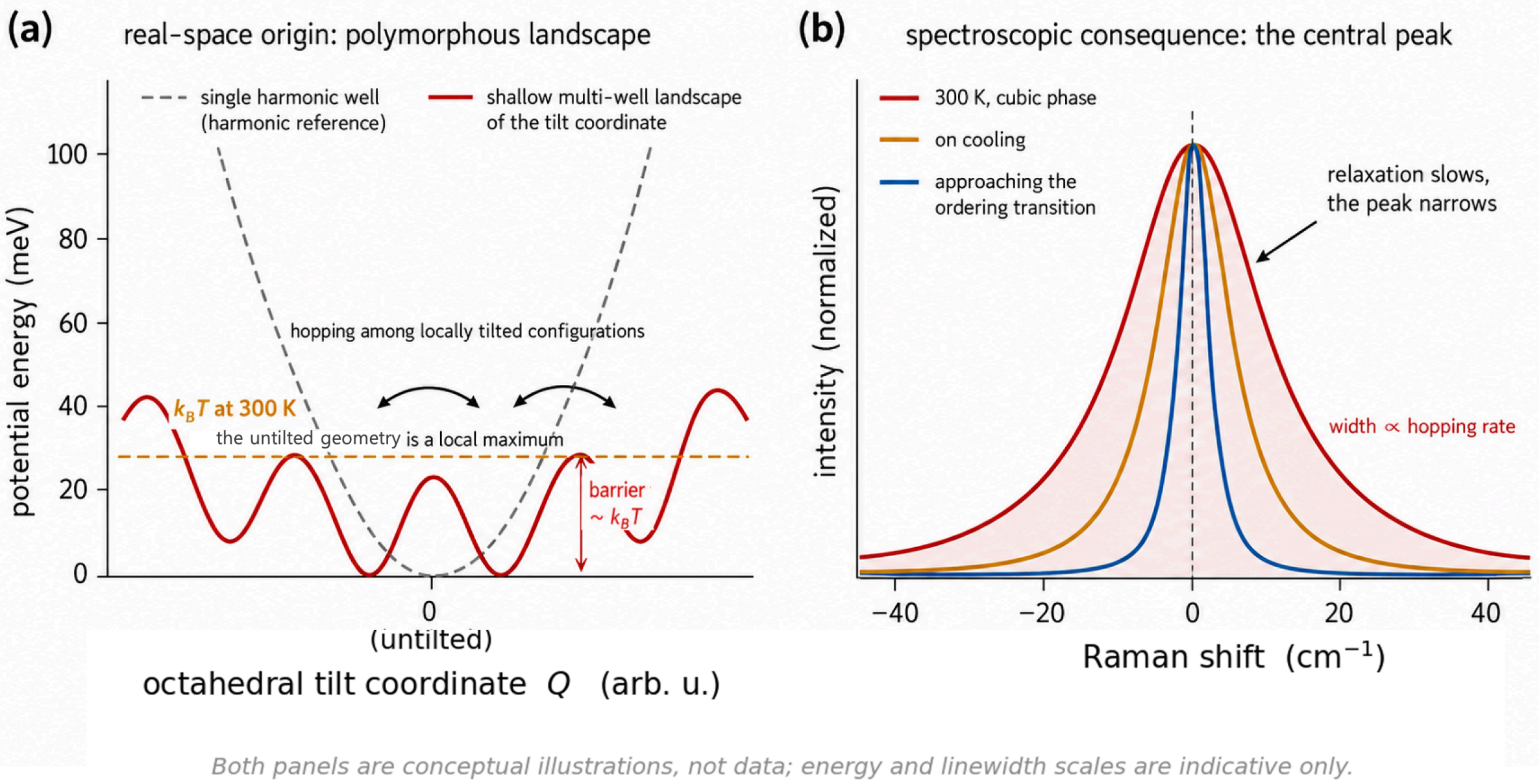

**Fig. 13 |** (a) The origin of the dynamics, schematically: along the octahedral tilt coordinate the potential is not a single harmonic well (dashed) but a shallow multi-well landscape (solid) whose barriers are comparable to $k_B T$ at room temperature, so that the system hops among locally tilted configurations rather than oscillating about an untilted one. This is the real-space content of the polymorphous description[21] of Section 3.5. (b) The spectroscopic consequence, schematically: a relaxational coordinate produces a quasielastic central peak whose width tracks the hopping rate; on cooling toward the ordering transition the relaxation slows and the peak narrows. Both panels are conceptual illustrations, not data[3].

"Figure adapted from [4], reproduced with permission in the final published version"

**Fig. 14 | The overdamped rotational branch in $CsPbBr_3$, energy-resolved:** inelastic-scattering spectra along the M-R zone edge showing linewidths exceeding the mode frequencies, the momentum-space companion to the diffuse-rods map of Figure 2. Reproduced with permission from ref. 4, Springer Nature Ltd.

The A-site cation adds a separate, genuinely molecular channel in the hybrids: the picosecond reorientational dynamics established in the 1980s (Section 2.2.1) and resolved structurally by neutron diffraction[29,30]. Two facts keep this channel in its proper place. It is absent in $CsPbBr_3$, which nonetheless shows the full anharmonic phenomenology; and it couples to the framework through hydrogen bonding (Section 3.3), so that in the hybrids the cation dynamics and the tilt dynamics are hybridized rather than independent. We treat the cation as a spectator that modifies rates and adds a dielectric channel (Section 6), not as the protagonist.

## 4.2. Thermal transport: the phonon-glass half of the duality

The thermal conductivity is where the softness of the lattice becomes a single number, through the elementary kinetic relation

$$\kappa_L = (1/3)\, C_v\, v_g\, \ell \tag{4}$$

relating $\kappa_L$ to the heat capacity per unit volume $C_v$, the phonon group velocity $v_g$, and the phonon mean free path $\ell$: once anharmonic scattering squeezes $\ell$ down to a few bond lengths, as it does throughout this family, $\kappa_L$ collapses toward the amorphous limit regardless of how structurally perfect the crystal is. Room-temperature lattice thermal conductivities are approximately 0.5 W $m^{-1}$ $K^{-1}$ in $MAPbI_3$ single crystals — with polycrystalline samples showing comparable values, indicating that the limit is intrinsic rather than microstructural — with resonant scattering of heat-carrying phonons by the rotational degrees of freedom identified as the dominant attenuation channel[43]. The all-inorganic compounds are equally poor conductors: 0.45, 0.42, and 0.38 W $m^{-1}$ $K^{-1}$ for single-crystalline $CsPbI_3$, $CsPbBr_3$, and $CsSnI_3$ nanowires respectively[44], values driven by low group velocities of the heavy, softly bonded framework and by extremely short phonon mean free paths — again independent of the organic cation. Single-crystal studies across the halide series add ultrahigh thermal expansion to the same picture[54]. These are among the lowest thermal conductivities known for fully dense stoichiometric crystals, in the range engineered deliberately in thermoelectrics; here the phonon-glass electron-crystal behaviour[14] is not engineered but constitutive. Two consequences propagate forward: hot-carrier cooling is throttled because the lattice cannot export heat (Section 7.4), and any transport theory that assumes a well-defined phonon lifetime for the low-lying optical modes starts from a fiction (Section 8).

## 4.3. What the harmonic framework gets right

It would misrepresent the field to suggest the harmonic picture is useless. Density-functional lattice dynamics in the orthorhombic phase reproduces measured infrared and Raman spectra well; the acoustic sector, elastic constants, and low-temperature heat capacities are conventionally described; and the soft-mode language of Section 3.3 correctly identifies *which* instabilities condense at the transitions — the $M_3^+$ and $R_4^+$ rotational modes — even where it misdescribes *how* they condense. The harmonic phonon basis also remains the reference frame in which anharmonicity is quantified: the statement "the M-R branch is overdamped" is a statement about harmonic eigenvectors with non-perturbative self-energies. The framework fails not by irrelevance but by the size of its corrections.

## 4.4. Where it fails: the anatomy of the anharmonicity

The failures are of three escalating kinds. (i) *Perturbative breakdown.* For the zone-edge rotational branches the imaginary part of the phonon self-energy exceeds the real frequency: the damped-oscillator lineshape is inadmissible and the response is relaxational, as in Figure 13(b). No resummation of phonon-phonon diagrams rescues a quasiparticle that completes less than one oscillation. (ii) *Reference-structure breakdown.* The harmonic expansion is taken about the cubic geometry, which Section 3.5 established is a saddle-like average, not a minimum: the true landscape is the multi-well of Figure 13(a), and the harmonic frequencies about the average structure are imaginary precisely for the modes that matter. Self-consistent phonon methods, which renormalize frequencies with thermal displacement amplitudes, repair the imaginary frequencies but not the multi-well kinetics. (iii) *Separability breakdown.* The expressions of anharmonicity are not mutually reducible: Cohen *et al.*[55] show that different halide perovskites express anharmonicity in qualitatively different ways — a warning against transferring a single anharmonic narrative across the family — and the two-dimensional correlations of Lanigan-Atkins *et al.*[4] show that the fluctuations are collective objects with their own emergent structure, not independent local hops.

For the electronic story to come, the operative summary is this: the lattice presents the carriers with a polarization environment whose fluctuations are (a) large — instantaneous local symmetry breaking comparable to the distortions separating phases; (b) slow — picosecond relaxational timescales overlapping the carrier scattering time; (c) spatially correlated — organized into quasi-two-dimensional textures; and (d) polar — the fluctuating coordinates carry dipole moments and modulate the band edges directly. Each of the four properties will be needed: (a) and (d) in the polymorphous electronic structure of Section 5, (b) in the dynamic screening of Section 6, (c) in the polaron structure of Section 7, and all four in the transport of Section 8.

### 4.5. Quantitative handles

Because "anharmonic" risks becoming a slogan, we record the handles by which it is measured. Spectroscopically: the ratio of linewidth to frequency for the rotational branches, and the central-peak fraction of the low-frequency Raman weight, both of order unity in the cubic phases[3,41,4]. Thermodynamically: the deviation of the heat capacity and thermal expansion from quasi-harmonic predictions[54], and the near-tricritical character of the tetragonal-cubic transition[30]. Computationally: the spread of the instantaneous band gap and local geometry in molecular dynamics, and the energy lowering of polymorphous relative to monomorphous descriptions[21,19,20,42]. These are the numbers a sceptic should ask for, and the primary literature supplies them.

### 4.6. Open questions

Three questions remain genuinely open. First, the microscopic identity of the central peak: relaxational tilt hopping, strongly overdamped soft modes, and coupled cation-cage dynamics produce similar quasielastic signatures, and their disentanglement — for instance by comparing polarization selection rules and momentum structure across the Cs/MA/FA series — is incomplete. Second, the dimensionality question: whether the two-dimensional rotational correlations of the cubic phase template two-dimensionally confined electronic states, as conjectured by Lanigan-Atkins *et al.*[4], is untested; it is among the sharpest falsifiable proposals in the field. Third, the transferability question raised by Cohen *et al.*[55]: a predictive theory should say *which* anharmonic expression a given composition will choose, and none currently does.

## 5. Electronic Structure

With the lattice in hand, we construct the electrons. The section proceeds from the chemistry of the band edges (5.1), through the relativistic physics that any quantitative treatment must include (5.2), to the spin textures that relativity makes possible (5.3), and finally to the correction that the lattice of Section 4 imposes on all of the above: the replacement of the band structure of a geometry by the band structure of a distribution (5.4-5.5).

### 5.1. The inverted band edges

The frontier orbitals were identified qualitatively by Weber[23] and quantitatively by the first-principles literature: the valence-band maximum of $APbX_3$ is the antibonding combination of the Pb 6s and halide np orbitals, and the conduction-band minimum is built from the Pb 6p states. Three structural facts of Section 3 translate directly into electronic ones. First, because the VBM is *antibonding*, its energy rises as the Pb-X overlap increases: compressing the lattice or straightening the Pb-X-Pb bridges by untilting the octahedra, closes the gap, while tilting opens it. The band gap is therefore a direct read-out of the tilt state, which is why the instantaneous gap fluctuates with the rotational dynamics of Section 4 and why the gap responds anomalously to temperature (Section 5.4). Second, both band edges are dominated by the framework: the direct gap sits at the R point of the cubic zone (folding to Gamma in the tilted phases), the dispersion near the extrema is large, and the effective masses are correspondingly light — the measured exciton reduced mass[1] of 0.104 $m_e$ implying electron and hole masses each of order 0.2 $m_e$. Third, the halide series tunes the gap through the np level: descending Cl to Br to I raises the VBM and closes the gap across the visible, the trend already visible in Weber's crystals. The near-symmetric, light electron and hole masses — a rarity among semiconductors, where holes are usually heavy — descend from the same inversion: the "s-like" band is here the *valence* band.

## 5.2. Relativity is not a correction

On lead, spin-orbit coupling enters at the electronvolt scale. Even *et al.*[38] established the magnitude and the consequence: the Pb 6p conduction manifold is split into a lower $j = 1/2$ doublet — which becomes the conduction-band minimum — and an upper $j = 3/2$ set roughly an electronvolt above, so that the true gap is the gap to the spin-orbit-stabilized $j = 1/2$ states. Neglecting SOC raises the calculated conduction band by about that electronvolt; in scalar-relativistic density-functional calculations this error is fortuitously cancelled by the usual gap underestimate, producing "correct" gaps from doubly wrong physics — the trap recorded in Section 2.4.1 and Appendix A. The clean treatment is a relativistic quasiparticle calculation; the quasiparticle self-consistent GW study of Brivio *et al.*[39] is the reference point, delivering gaps, dispersions, and the strongly spin-orbit-coupled character of both band edges without empirical adjustment. An immediate spectroscopic consequence of the $j = 1/2$ conduction character is the anomalously simple band-edge optics: the transition is between two doubly degenerate bands with near-isotropic dipole coupling, which is part of why the effective-mass and hydrogenic-exciton analyses of Section 7 work as well as they do — and, in nanocrystals, why the exciton fine structure takes the unusual[56] form discussed in Section 10.3.

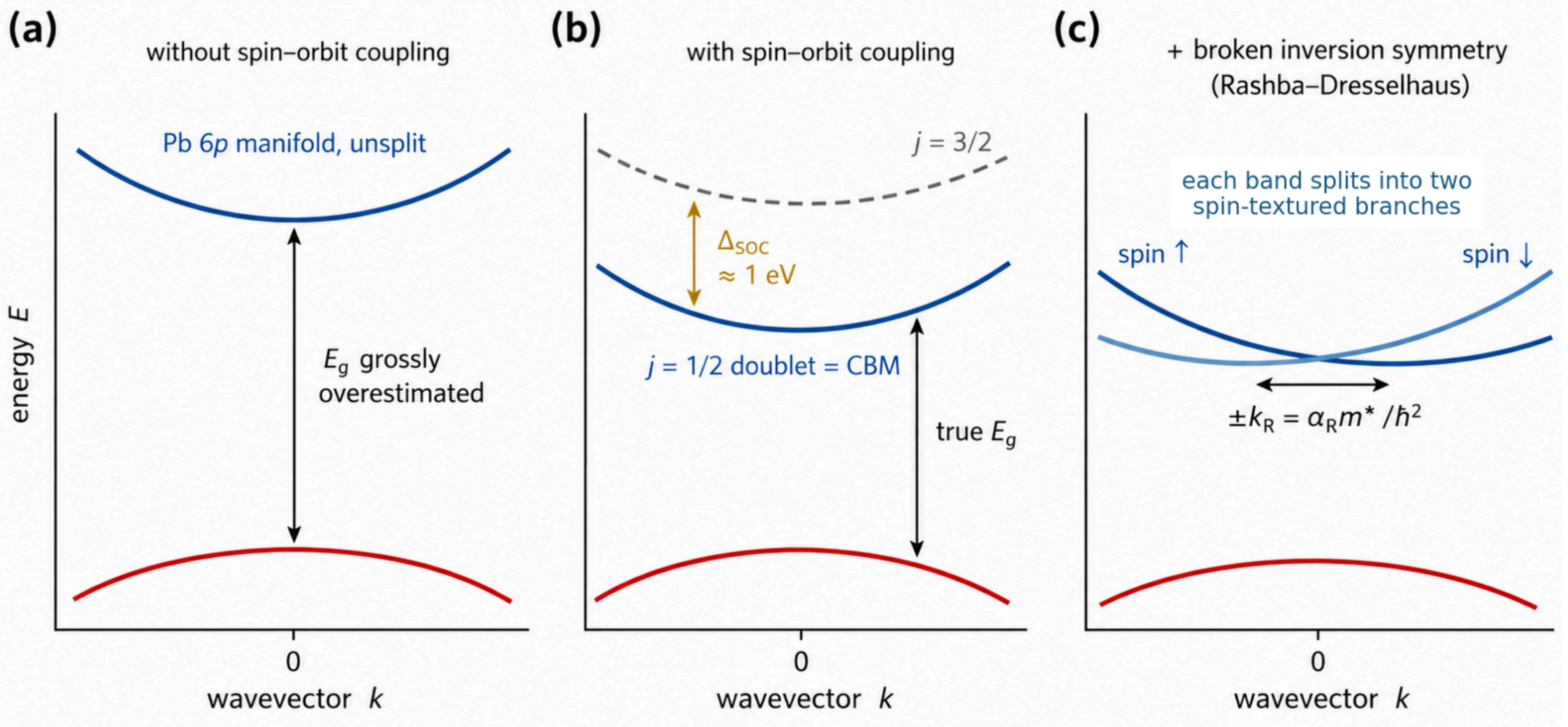

**Fig. 15 | The relativistic construction of the band edges, schematically.** (a) Without spin-orbit coupling the gap is grossly overestimated at fixed exchange-correlation treatment. (b) SOC splits the Pb 6p manifold, lowering the $j = 1/2$ doublet by of order an electronvolt to form the true conduction-band minimum[38,39]. (c) If inversion symmetry is broken — statically in polar phases or transiently by the fluctuations of Section 4 — the combination of strong SOC and the polar field splits each band into two spin-textured branches displaced by ±kR (Rashba-Dresselhaus effect[57,17]). Energy and momentum scales are illustrative only.

## 5.3. Rashba-Dresselhaus physics: real, but conditional

Strong SOC plus broken inversion symmetry yields spin-momentum-locked band splitting of the generic Rashba form

$$E_{\pm}(k) = \hbar^2 k^2 / (2m^*) \pm \alpha_R k \quad (8)$$

where $\alpha_R$ is the Rashba parameter and the two spin-split branches are displaced in momentum by $\pm k_R = \alpha_R m^*/\hbar^2$. Both ingredients — strong SOC and broken inversion symmetry — are available here: the first unconditionally, the second conditionally. In a structure whose space group is centrosymmetric, no static Rashba splitting exists; the question is therefore *whether and where* inversion symmetry is broken. Three answers coexist in the literature, and they do not agree. In genuinely non-centrosymmetric configurations — polar distortions, surfaces and interfaces, some layered compounds — static splittings appear in relativistic calculations at the band extrema[39,17]. In nominally centrosymmetric bulk phases, the local symmetry breaking of the polymorphous, dynamically disordered lattice (Sections 3.5, 4) has been proposed to generate transient, fluctuating Rashba fields — a "dynamical Rashba" effect whose observable consequences (spin relaxation channels, weak signatures in nonlinear optics) would differ from those of a static splitting. This second claim is directly contested: second-harmonic-generation rotational-anisotropy measurements on $MAPbI_3$ single crystals find the tetragonal structure centrosymmetric (I4/mcm) within experimental resolution and report no signature of either a static or a dynamic bulk Rashba effect, attributing earlier large computed splittings to unrelaxed, artificially polar structural models[92]. A proposed consequence of either variant — an indirect-in-*k* character of the effective gap, with spin-split extrema displaced in momentum, suppressing radiative recombination — has been advanced as a contributor to the long carrier lifetimes, but remains a qualitative proposal rather than a quantitatively worked mechanism. We record our position: the existence of Rashba physics in genuinely non-centrosymmetric configurations is settled; a bulk dynamical Rashba effect in the nominally cubic and tetragonal phases is not established and is directly challenged by the centrosymmetry measurement of Frohna *et al.*[92]; and its relevance to photovoltaic carrier lifetimes should currently be read as a hypothesis rather than a demonstrated mechanism. The spintronic ramifications are collected in Section 11.

## 5.4. The band structure of a distribution

Section 4 obliges a reinterpretation of everything above. The band structure of "cubic $APbX_3$" computed on the ideal Pm-3m geometry is the band structure of a configuration the crystal never occupies. Zhao *et al.*[21] quantify the correction: allowing the nominally cubic cell to relax into its thermal distribution of locally tilted polymorphous configurations *opens* the gap substantially relative to the ideal geometry, modifies the effective masses, and generates local spin splittings even where the average structure is centrosymmetric. Molecular-dynamics studies add the time domain: the instantaneous gap fluctuates by an appreciable fraction of its mean on few-hundred-femtosecond timescales, driven by the same rotational and polar coordinates that produce the central peak[19,20]. Two experimental anomalies fall out naturally. The temperature coefficient of the gap is *positive* — the gap opens on heating — opposite to conventional semiconductors: thermal population of the tilt fluctuations bends the Pb-X-Pb bridges and, through the antibonding-VBM mechanism of Section 5.1, pushes the valence band down faster than thermal expansion closes the gap. And Urbach tails remain steep despite the violent gap fluctuation, because the fluctuations are fast and spatially correlated rather than static and random — the carriers average over them, a motional narrowing whose full treatment belongs to the polaron section (Section 7).

See Figure 12 (Section 3.5) for the corresponding first-principles comparison of monomorphous and polymorphous band gaps, densities of states, and instantaneous band-gap fluctuation.

### 5.5. Status of the framework

The four-part verdict for this section: the experimental facts (gaps, masses, optical selection rules, halide trends) map onto a relativistic band framework with unusual cleanliness; the framework's standard *practice* — computing on the average geometry — fails at the tens-of-per-cent level for the gap and qualitatively for symmetry-derived quantities; the repaired practice — configurational averaging over the polymorphous ensemble — restores quantitative contact[21,58]; and the open question is dynamical: what replaces the Bloch quasiparticle when the averaging timescale and the scattering timescale merge. That question is Section 7's.

## 6. Dielectric Response and Screening

The dielectric function is the hinge between the lattice of Section 4 and the quasiparticles of Section 7: every Coulomb interaction in the problem — electron-hole binding, carrier-defect scattering, carrier-carrier repulsion — is screened by ε, and in this family ε depends violently on the frequency at which one asks. The structure of this section is the structure of ε(ω) itself, descending from optical to static frequencies and assigning each step to a lattice degree of freedom established earlier.

### 6.1. The staircase

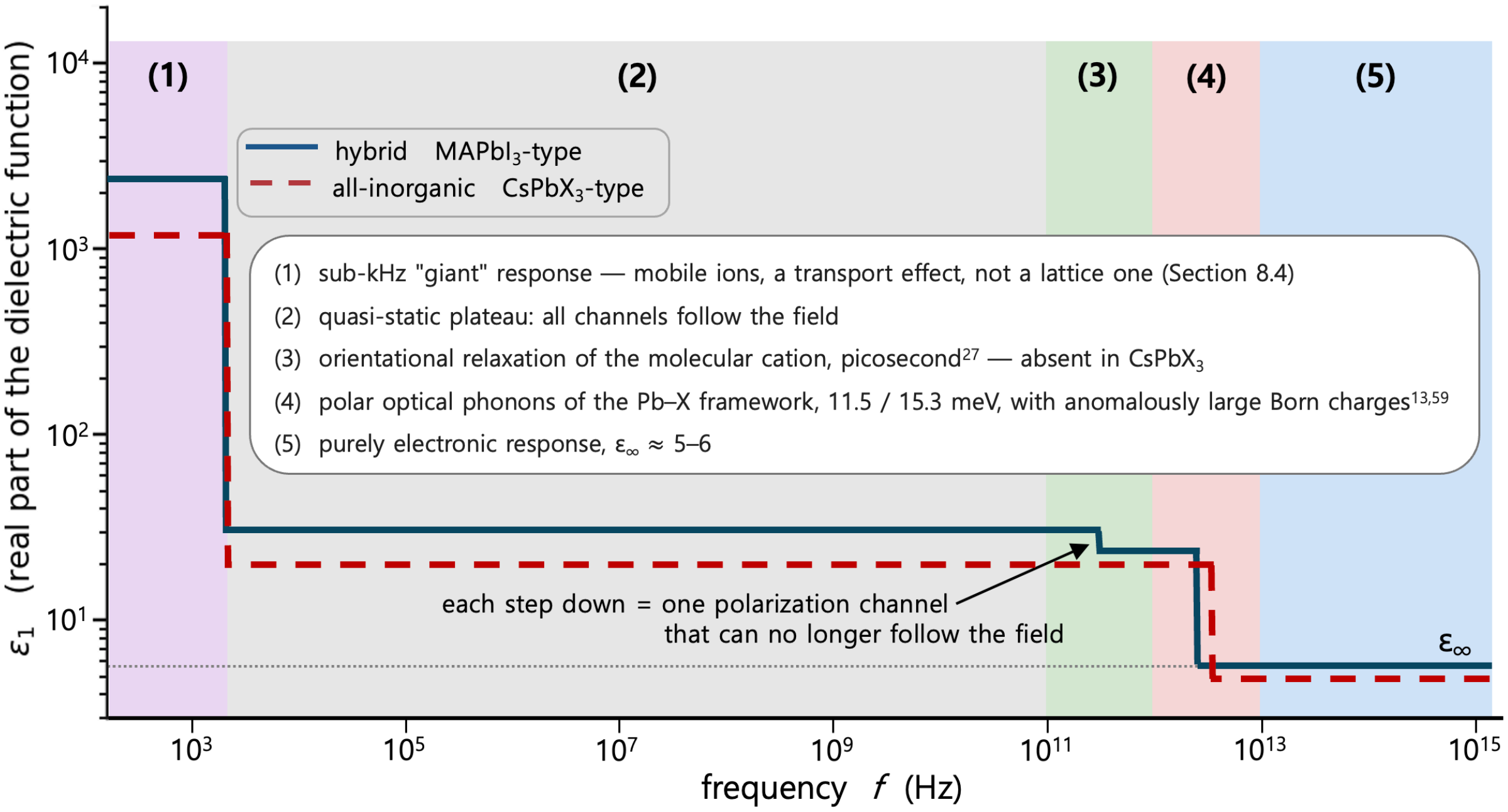

**Fig. 16 | The real part of the dielectric function, schematically, for a hybrid (solid) and an all-inorganic (dashed) lead-halide perovskite.** Each step downward with increasing frequency marks a polarization channel that can no longer follow the field: ionic space-charge effects in the sub-kHz "giant" response (a transport phenomenon, Section 8.4, not a lattice one); the

orientational relaxation of the molecular cation in the hybrids, at picosecond timescales[27]; the polar optical phonons of the Pb-X framework in the terahertz range[13,59]; and finally the purely electronic response $\varepsilon_\infty$. Axis values are indicative only; measured spectra should be taken from the cited sources.

Reading the staircase from the right: the electronic (optical) dielectric constant is modest, of order 5-6 across the family — these are not high-refractive-index semiconductors — so the *unscreened* electron-hole interaction would be strong. The first great enhancement is ionic: the polar Pb-X stretching and bending modes, at the 11.5 and 15.3 meV energies quantified by Wright *et al.*[13] and characterized across the halides by Sendner *et al.*[59], carry anomalously large Born effective charges (the dynamical charges exceed the formal ones — a signature of the cross-gap Pb 6s - X np hybridization of Section 5.1, whereby bond stretching transfers charge across the bond), and raise the quasi-static lattice response to several times the electronic value. In the hybrids a further orientational step follows at lower frequency: the picosecond dipolar relaxation of the MA cation resolved by Poglitsch and Weber[27], absent by construction in $CsPbX_3$ — the cleanest single illustration in the field of which dielectric physics is molecular and which is framework. Below kilohertz frequencies the measured response rises again by orders of magnitude; this "giant" low-frequency response is not a polarization of the lattice but the signature of mobile ions rearranging[60] and belongs to Section 8.

## 6.2. Which ε does a carrier see?

The staircase converts a static question into a kinematic one: a charge probing the lattice at frequency $\omega$ — equivalently, on timescale $1/\omega$ — is screened by $\varepsilon$ at that frequency. An electron and hole orbiting each other with a binding energy of tens of meV circulate at terahertz frequencies, *within* the phonon step: they are screened by more than $\varepsilon_\infty$ but less than the full static response, which is precisely why exciton binding energies extracted with either limiting $\varepsilon$ scattered across an order of magnitude before the model-independent measurement settled the matter (Section 2.2.2)[1,32]. A slow drifting carrier, by contrast, is dressed by the full lattice response — the polaron of Section 7. And a hot carrier, in the first hundreds of femtoseconds after photoexcitation, is initially screened almost electronically and then watches the lattice polarization build around it in real time: the time-domain observation of exactly this build-up — sub-picosecond dynamic screening protecting energetic carriers — is the content of Zhu *et al.*[11], and its emission-side counterpart, a time-dependent Stokes shift with the character of solvation in a polar liquid, is reported by Guo *et al.*[61], who name it *dielectric solvation*. The dielectric function of this material is thus better read as a solvation spectrum than as a constant, and the language of liquid-state chemistry — reorganization energy, solvation time — transfers with surprisingly little friction. That transfer is the quantitative face of the crystal-liquid duality.

## 6.3. Screening of defects, and a first look forward

The same staircase acts on carrier-defect Coulomb interactions. A charged defect at rest is screened by the full quasi-static response — several-fold stronger than in a conventional semiconductor of similar gap — which shrinks capture cross sections and shallow-level binding energies alike; a carrier passing quickly feels less. The screening contribution to the benignity of defects is therefore real but frequency-weighted, and its proper evaluation requires the polaronic treatment rather than a static $\varepsilon$ inserted into hydrogenic formulas. We defer the quantitative dispute to Section 9, noting here only the structural point: the same physics — soft polar lattice, large dynamical charges, slow relaxational channels — that sets the exciton scale (Section 7.1), dresses the carrier (Section 7.2–7.3), and throttles cooling (Section 7.4) also stands between every carrier and every charged defect. This provides the dielectric component of the coupled framework proposed in Section 1.2.

### 6.4. Open questions

First, the momentum dependence: the staircase above is the q -> 0 response, but the correlated quasi-two-dimensional fluctuations of Section 4 imply structured screening at finite q, essentially unexplored experimentally. Second, the nonlinear regime: at the carrier densities of operating devices and transient-absorption experiments, the polarization channels saturate and cross-couple (Auger heating and hot-phonon effects, Section 7.4), and no consistent density-dependent ε(ω, q, n) exists even empirically.

# 7. Quasiparticles: Excitons and Polarons

This section assembles the objects that carry the physics: the exciton (7.1), the large polaron (7.2), the anharmonically dressed hybrid of the two (7.3), and the hot carrier whose relaxation exposes the machinery in the time domain (7.4). Section 7.5 then discharges the promise of Section 1.2: it states the case that the three explanatory traditions of the field are one, and where the case is still open.

### 7.1. The exciton

The excitonic literature of this family has been surveyed in detail by Baranowski and Plochocka[62]; we extract here only what the later sections require. The bound electron-hole state of the three-dimensional compounds is a textbook Wannier-Mott exciton — a hydrogenic series whose binding energy scales with the reduced mass and the screening dielectric constant exactly as the hydrogen Rydberg does:

$$E_b = (\mu / m_0)\,(1 / \varepsilon_r^2) \times 13.6\ \text{eV} \qquad (5)$$

with one non-textbook feature: its binding energy sits in the crossfire of the dielectric staircase of Section 6. The settled numbers, from magneto-absorption in pulsed fields up to 150 T resolving excited states and the free-carrier Landau fan[1,32]: reduced mass 0.104 ± 0.003 $m_e$ and binding energy near 16 meV in the low-temperature phase of $MAPbI_3$, decreasing by several meV in the room-temperature phase, with the bromides binding more strongly in proportion to their wider gaps and stiffer screening. Two readings follow. Practically: at room temperature the binding energy is comparable to $k_BT$, so photoexcitation yields predominantly free carriers — the fact that makes these materials photovoltaic absorbers rather than excitonic ones, and the single most consequential number in the family's device physics. Fundamentally: the *effective* dielectric constant required to reconcile mass and binding energy lies between the optical and static limits, confirming that the exciton is a dynamically screened object whose internal motion competes with the phonon response — a polaronic exciton avant la lettre, and the reason the naive hydrogenic formula with either limiting ε fails (Sections 2.2.2, 6.2). The 1990s magneto-optics that anticipated all of this[22,31] deserves its restitution here.

### 7.2. The Fröhlich large polaron

A slow carrier in a polar lattice permanently drags a cloud of lattice polarization; carrier plus cloud is the Fröhlich polaron[63], characterized by the dimensionless coupling constant α built from the LO phonon frequency and the difference of inverse dielectric constants. With the measured LO energies and dielectric steps, the lead halides land at α of roughly 1.7-3 across the family — squarely in the *intermediate* regime: too strong for lowest-order perturbation theory, far too weak for self-trapping, and precisely where the Feynman variational treatment[64] is the tool of choice. The predicted phenomenology at intermediate coupling — mass enhancement of tens of per cent, binding of order the LO quantum, radius of several nanometres spanning many unit cells — is what distinguishes the *large* polaron, which remains a mobile band-like object, from the small polaron's site-localized hopper; nothing in the transport of Section 8 suggests the latter. First-principles many-body theory concurs: multiphonon Fröhlich calculations for $MAPbI_3$ yield polaronic mass enhancement and carrier lifetimes consistent with the intermediate-coupling assignment[40], and path-integral mobility calculations on the Feynman model reproduce the measured mobility scale with no free parameters[65]. Within the general taxonomy of polarons in materials[16], the lead halides are arguably the cleanest experimentally accessible realization of the intermediate-coupling Fröhlich problem — the regime the formalism was built for, and materials rarely occupy.

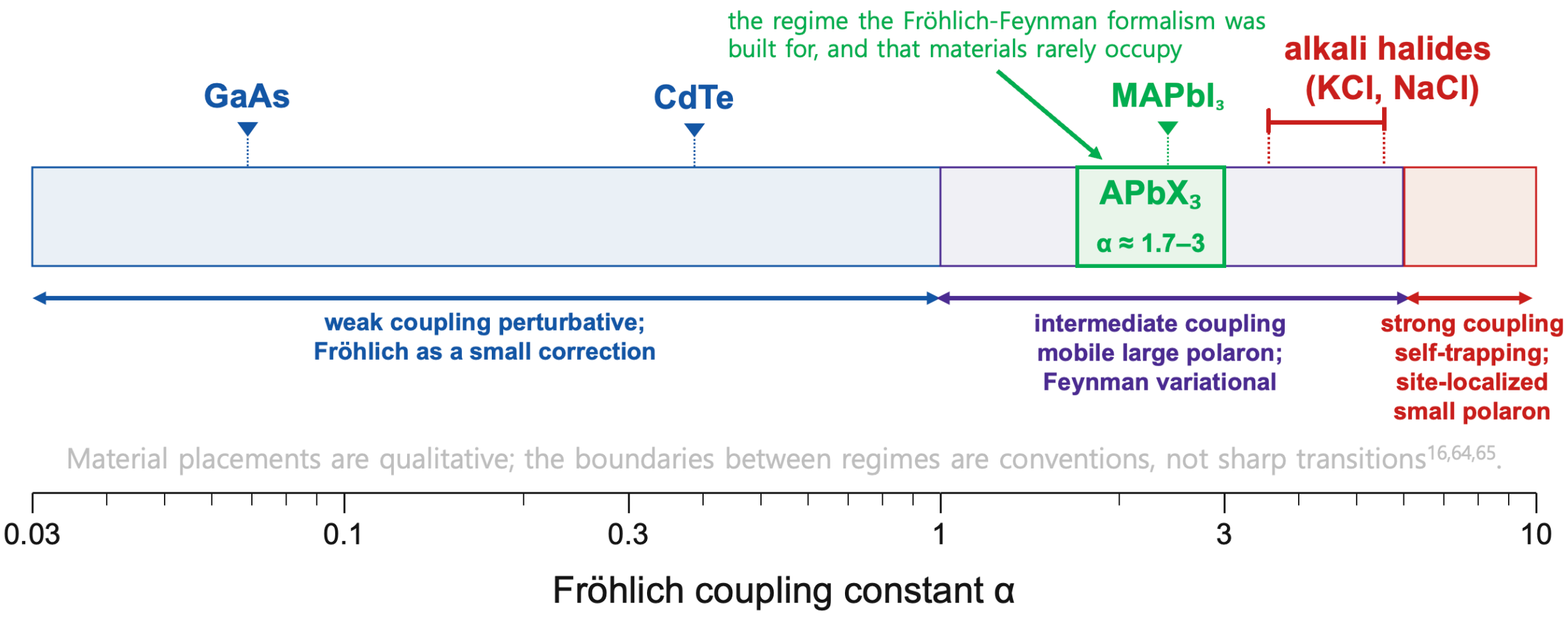

**Fig. 17 | The Fröhlich coupling landscape, schematically.** Weak-coupling semiconductors (GaAs) are perturbative; strong-coupling ionic insulators (alkali halides) self-trap; the lead-halide perovskites occupy the intermediate window $\alpha \approx 1.7$-3 where the polaron is a heavy but mobile large polaron and the Feynman variational description applies[16,64,65]. Material placements are qualitative.

## 7.3. Beyond textbook Fröhlich: the anharmonic dressing

The textbook Fröhlich Hamiltonian assumes harmonic phonons with well-defined frequencies. Section 4 removed that assumption, and the experiments show what replaces it. Time-resolved spectroscopy finds polaron formation — the build-up of the dressing — completing on sub-picosecond to picosecond timescales in $MAPbBr_3$ and $CsPbBr_3$ alike[12], and the dressing medium is not a phonon bath but the relaxational, liquid-like polarization of Sections 4 and 6: the dynamic-screening observation of Zhu *et al.*[11] and the dielectric-solvation Stokes shift of Guo *et al.*[61] are polaron formation watched from the carrier's side and the lattice's side respectively. Theory has followed the experiments into this regime: treating the carrier in a lattice whose fluctuations are large, slow, and correlated — rather than as a perturbative phonon gas — reproduces the transport phenomenology and reframes the polaron as a carrier surfing dynamic disorder[10,19,20,42,66]. The proposal of Lanigan-Atkins *et al.*[4], that the quasi-two-dimensional rotational textures template correspondingly anisotropic polarons, remains the most specific untested prediction in this area (Section 4.6). The exciton inherits the same physics: at room temperature the electron-hole pair is a correlated pair of dressed carriers — an exciton-polaron — whose binding, screening, and lifetime are inseparable[67], which is the precise sense in which the binding-energy controversy of Section 2.2.2 was never a measurement dispute but a definition dispute.

## 7.4. Hot carriers: the machinery in the time domain

Above-gap excitation launches carriers that must shed their excess energy through the same channels, and their cooling is anomalously slow — by orders of magnitude at high density relative to GaAs-class semiconductors. The observations: a pronounced hot-phonon bottleneck in lead iodide perovskites at elevated carrier density[69]; fluence-resolved kinetics in $MAPbI_3$ showing Fröhlich-mediated cooling through zone-centre LO emission with an LO lifetime near 0.6 ps, bottlenecked by suppression of the Klemens decay channel at densities around $10^{18}$ $cm^{-3}$ and further slowed by Auger reheating[70] around $10^{19}$ $cm^{-3}$; and acoustic-to-optical phonon up-conversion sustaining the hot LO population, markedly stronger in the hybrids — $FAPbI_3$ cooling an order of magnitude more slowly than the caesium analogue — because the organic cation's dense low-frequency spectrum brokers the up-conversion and the phonon-glass thermal transport (Section 4.2) prevents the heat from leaving[71]. Polaron formation itself contributes at low density: energy deposited into the local dressing relaxes on the solvation timescale rather than the bare-phonon one (Zhu *et al.*, 2016; a polaronic cooling model in these terms is developed by Frost, Whalley, and Walsh[68]). The mechanistic decomposition remains partly contested, but every proposed ingredient — Fröhlich dominance, hot phonons, up-conversion, glassy thermal transport, polaronic solvation — is a Section 4–6 ingredient. Hot-carrier physics is not a separate subject; it is the same subject, strobed.

## 7.5. The unification, defended and delimited

Section 1.2 promised the argument; here it is, in four steps. (1) The Pb $6s^2$ lone pair and the soft polar Pb–X bond jointly influences the electronic structure (antibonding VBM, strong SOC: Section 5) and the lattice character (shallow multi-well landscape, large Born charges and low-lying polar modes: Sections 3–4). These can be viewed as related consequences of the same bond chemistry, expressed in different sectors. (2) This chemistry contributes to the dielectric staircase of Section 6, whose steps correspond to distinct polarization channels. (3) A carrier or exciton immersed in this environment is a dynamically screened, dressed object — an intermediate-coupling, anharmonically dressed polaron in the framework of Sections 7.2–7.3 — with measured masses, binding energies, formation times and cooling anomalies reproduced quantitatively in cases where they have been calculated. (4) Transport and some aspects of defect behaviour (Sections 8–9) can then be interpreted in terms of this dressed object. The three explanatory traditions of Section 1.2 — electronic, dielectric-polaronic and dynamical — are therefore best viewed as coupled consequences of the same soft polar framework rather than as a single microscopic mechanism. The delimitation is equally important: this framework is only partly sufficient for defects. Whether dressing and screening explain a substantial fraction of nonradiative tolerance, or whether defect chemistry contributes independently, remains the live dispute of Section 9.

# 8. Charge Transport

Transport is where the dressed quasiparticle of Section 7 meets the measurement, and where the field's most persistent quantitative puzzle lives. The facts first (8.1), the framework and the controversy second (8.2-8.3), the ionic channel third (8.4).

## 8.1. The facts

Room-temperature mobilities cluster in the tens of $cm^2V^{-1}s^{-1}$ for good single crystals and somewhat below for films, remarkably insensitive to the enormous variation in preparation and defect content — the first hint that the limit is intrinsic[2,72]. Hall and photoconductivity measurements resolve long recombination lifetimes coexisting with these modest mobilities[73], and single crystals show trap densities extraordinarily low for solution-grown material[74], giving the micrometre-class diffusion lengths[6] of Section 2.3. The temperature dependence is the signature fact, conventionally parametrized as a power law

$$\mu(T) = \mu_0 \, (T/T_0)^{-n} \tag{10}$$

with the textbook acoustic-deformation-potential value $n = 3/2$; the mobility of $APbX_3$ rises on cooling with reported exponents spanning roughly 1.3-1.6 depending on sample and temperature window — close to, but not cleanly pinned on, the textbook value. The combination — modest magnitude, intrinsic character, an approximately $T^{-3/2}$ law — is the explanandum.

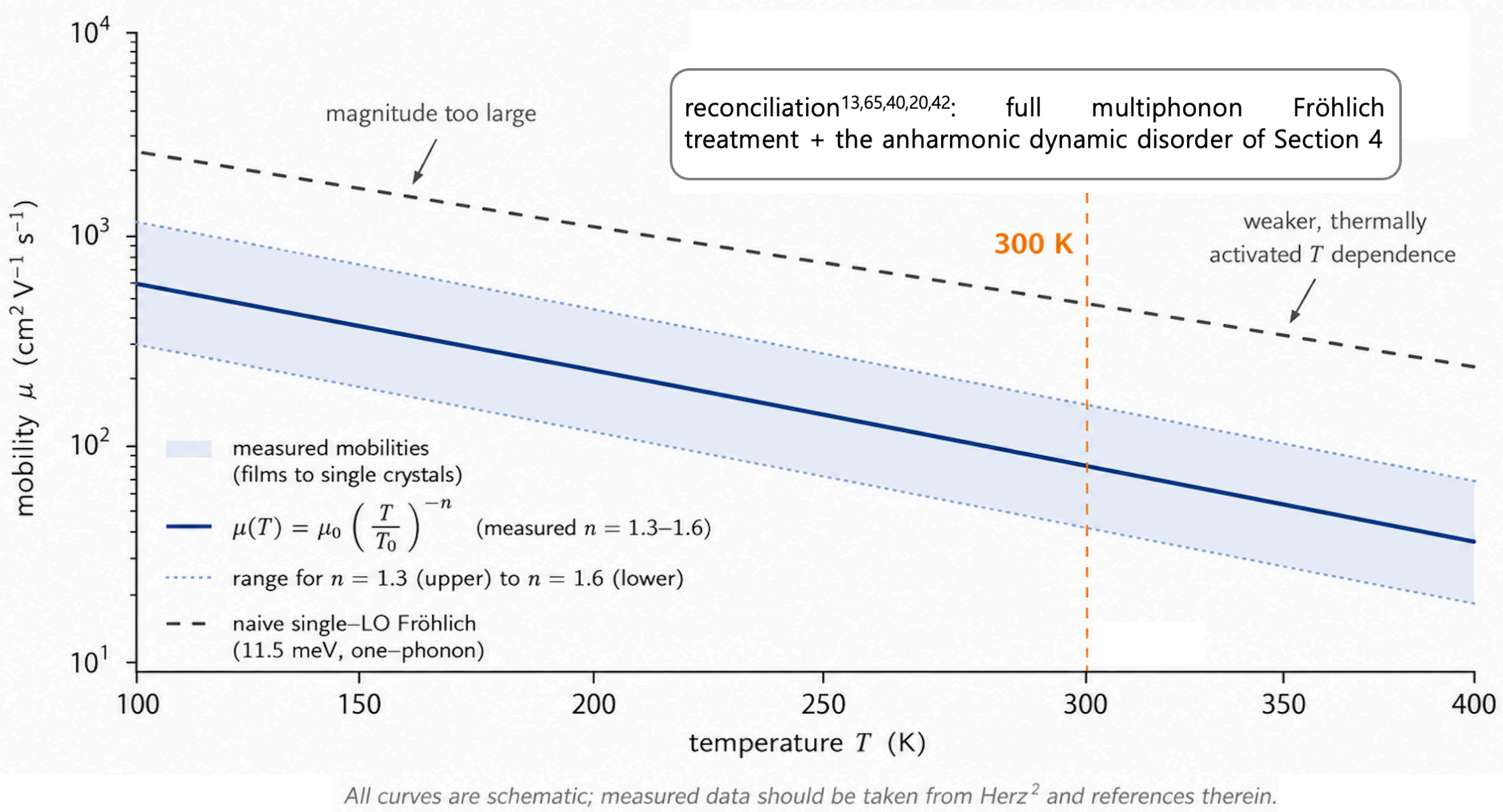

**Fig. 18 | The transport puzzle, schematically.** Measured mobilities (band, indicative of the spread between films and crystals) rise on cooling roughly as $T^{-3/2}$; a single-LO-phonon Fröhlich calculation (dashed) predicts both a different magnitude and a weaker, thermally activated temperature dependence in the naive treatment. The reconciliation — full multiphonon Fröhlich treatment plus the anharmonic dynamic disorder of Section 4 — is discussed in the text[13,20,40,42,65]. All curves are schematic; measured data should be taken from Herz[2] and references therein.

## 8.2. The intrinsic limit: Fröhlich scattering, properly done

The scattering-mechanism question was settled spectroscopically before it was settled theoretically: the temperature-dependent emission linewidths of Wright *et al.*[13] show Fröhlich coupling to the low-lying LO modes dominating at room temperature, with acoustic (deformation-potential) scattering negligible and impurity scattering visible only at low temperature — ruling out at a stroke the defect-limited interpretation of the modest mobilities. The quantitative implementations follow two roads that converge. The Feynman path-integral route treats the measured α and LO frequencies variationally and delivers room-temperature mobilities on the measured scale with no adjustable parameters[65]. The ab initio route computes the electron-phonon vertex from first principles and finds that *multiphonon* processes are essential — the coupling is strong enough that one-phonon golden-rule rates are wrong not by refinement but by category[40]. Both roads agree on the headline: the mobility of $APbX_3$ is intrinsically capped at the tens-to-low-hundreds of $cm^2\ V^{-1}\ s^{-1}$ scales by the same polar softness that produces every other anomaly in this review. There is no defect story to tell about the magnitude, and cleaner samples will not transform it[2].

## 8.3. The $T^{-3/2}$ controversy and the dynamic-disorder resolution

The temperature dependence is harder to explain than the magnitude. Experimentally, the mobility follows an approximately $T^{-3/2}$ dependence over the reported temperature range. A $T^{-3/2}$ law is classically associated with acoustic deformation-potential scattering in a band semiconductor; however, the temperature-dependent linewidth data indicate that acoustic scattering is not the dominant room-temperature scattering channel. Conventional Fröhlich scattering by a 15 meV LO mode predicts a weaker and partly activated temperature dependence. Several mechanisms could therefore contribute to the observed temperature dependence and are not mutually exclusive: multiphonon processes may reshape the temperature law[40]; the temperature dependence of the polaron itself may enter through the Feynman mobility[65]; and transport through the slow, correlated disorder field of Section 4 may generate a similarly steep negative temperature dependence. Explicit treatments of transport through correlated anharmonic disorder reproduce both the mobility scale and a steep negative temperature dependence and identify the relevant correlations — which fluctuations couple and over what range — as important controlling quantities[19,20,42,66]. On this basis, the $T^{-3/2}$ exponent need not be interpreted as a unique fingerprint of a single scattering mechanism. It may instead represent an emergent slope within a crossover regime in which carrier–phonon coupling, polaronic dressing and correlated dynamic disorder contribute simultaneously. This interpretation provides a physically plausible alternative to assigning the exponent directly to textbook acoustic scattering, but it remains a hypothesis rather than a uniquely established microscopic mechanism. Distinguishing “dressed carrier scattered by residual fluctuations” from “carrier of a fluctuating band structure” is therefore an important open problem. Measurements of the Hall factor, magnetotransport and frequency-resolved conductivity across the solvation window could help discriminate between these pictures.

## 8.4. The second conductor: ions

Underneath the electronic channel runs an ionic one. First-principles migration barriers identify vacancy-assisted iodide hopping as facile — activation energy near 0.6 eV, in agreement with kinetic data extracted from device transients — with the cation channels (MA at higher barrier, Pb effectively immobile) subordinate: the material is a mixed ionic-electronic conductor at operating temperature[60]. The consequences ramify through every measurement class in this review: the "giant" sub-kHz dielectric response of Section 6.1, current-voltage hysteresis, slow photoconductivity transients, field-driven halide segregation, and — because a migrating vacancy is a migrating *defect* — a defect population that is mobile, self-rearranging, and capable of the healing phenomenology taken up in Section 9. Methodologically the lesson is severe: any electronic-transport measurement slower than the ionic response times is contaminated by ionic rearrangement, and part of the early literature's scatter is attributable to exactly this. The clean electronic numbers of Section 8.1 are clean because their techniques (THz and microwave conductivity, time-resolved spectroscopy, fast Hall) outrun the ions.

## 8.5. Status

Facts mapped: intrinsic Fröhlich-limited magnitude — framework succeeds quantitatively. Framework strained: the temperature law, where the phonon-gas language gives way to dynamic disorder. Framework abandoned: any picture ignoring the ionic channel. Open: the observables that would close Section 8.3, and the finite-frequency conductivity across the dielectric staircase, which no experiment has yet traced continuously.

**Table 2.** Representative physical parameters discussed in the text, collected for reference. Family-wide entries are order-of-magnitude ranges rather than single numbers established for every compound; see the cited sections for the underlying measurements and their caveats.

| Quantity | Value | Compound(s) | Section / Refs |
|---|---|---|---|
| Exciton reduced mass | $0.104 \pm 0.003\ m_0$ | $MAPbI_3$ (low-T) | Sec. 7.1 / [1,32] |
| Exciton binding energy | ≈16 meV (low-T) | $MAPbI_3$ | Sec. 7.1 / [1,32] |
| LO phonon energy | 11.5 meV | $MAPbI_3$ | Sec. 6.1, Sec. 7.2 / [13] |
| LO phonon energy | 15.3 meV | $MAPbBr_3$ | Sec. 6.1, Sec. 7.2 / [13] |
| Fröhlich coupling constant α | ≈1.7–3 | family-wide | Sec. 7.2 / [13,16,64,65] |
| Room-temperature mobility | tens of $cm^2\ V^{-1}\ s^{-1}$ | single crystals, family-wide | Sec. 8.1 / [2,72] |
| Mobility temperature exponent $n$ | 1.3–1.6 (measured); 3/2 (textbook) | family-wide | Sec. 8.1, Sec. 8.3, Eq. 10 |
| Lattice thermal conductivity $\kappa_L$ | ≈0.5 $W\ m^{-1}\ K^{-1}$ | $MAPbI_3$, 300 K | Sec. 4.2 / [43] |
| Lattice thermal conductivity $\kappa_L$ | 0.45 / 0.42 / 0.38 $W\ m^{-1}\ K^{-1}$ | $CsPbI_3$ / $CsPbBr_3$ / $CsSnI_3$ nanowires, 300 K | Sec. 4.2 / [44] |
| Pb 6$p$ spin-orbit splitting | ≈1 eV | family-wide (Pb-based) | Sec. 5.2 / [38] |

# 9. Defects and Disorder

This is the section in which the review's organizing hypothesis meets its hardest test, and we have promised (Sections 1.2, 7.5) not to let the hypothesis win by assertion. The structure: the claim and its electronic basis (9.1), the supporting evidence (9.2), the counter-case (9.3), the screening and dynamics that complicate both (9.4), self-healing and the mobile defect population (9.5), and an explicit adjudication (9.6).

## 9.1. The defect-tolerance claim

The claim, in its strong form: the native point defects of $APbX_3$, though abundant, are electronically benign, because the low-formation-energy defects produce only shallow or resonant states while the deep-state-producing defects are too costly to form in numbers. The electronic basis is the band-edge inversion of Section 5.1, displayed in Figure 19: where a bonding VBM (GaAs-like) converts broken bonds into mid-gap traps, the antibonding Pb 6s - X np VBM pushes the corresponding dangling-bond states up toward or into the valence band, while the SOC-lowered conduction band descends to meet, from above, the states that would otherwise fall deep. The founding calculation is Yin, Shi, and Yan[7] for $MAPbI_3$ — dominant low-cost defects shallow, deep centres costly — with the same architecture identified in $CsPbBr_3$,[76] demonstrating that the tolerance, if real, is a framework property. Walsh and Zunger[75] systematized the argument into design rules, and in doing so made its logical structure explicit: defect tolerance is a *conjunction* of an electronic-structure condition (where levels sit) and a thermodynamic condition (which defects form), either of which can fail independently.

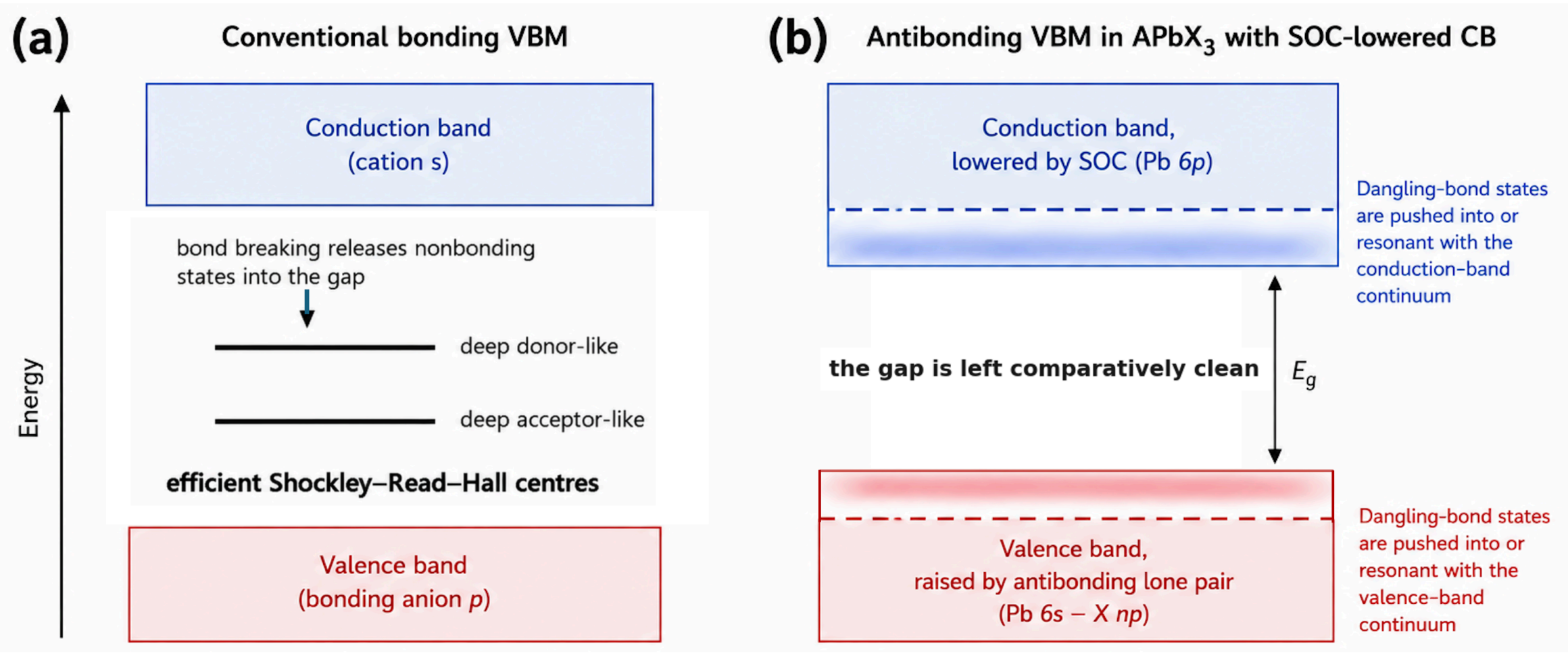

**Fig. 19 | The electronic-structure half of the defect-tolerance argument, schematically[7,75].** (a) A conventional bonding VBM: bond-breaking releases nonbonding states into the gap, where they are efficient Shockley-Read-Hall centres. (b) The antibonding VBM of $APbX_3$ with SOC-lowered conduction band: the corresponding states are pushed toward or into the continua, leaving the gap comparatively clean. The diagram states the *hypothesis*; its quantitative adequacy is contested in the text (Section 9.3).

## 9.2. The supporting evidence

Circumstantially, the case is strong. Solution-grown crystals achieve trap densities that would be respectable for zone-refined semiconductors[74]; lifetimes and diffusion lengths[6,73] coexist with chemical defect densities that should, on conventional Shockley-Read-Hall accounting with ordinary capture cross sections, be lethal; and the phenomenology is robust across preparation routes to a degree unknown in conventional semiconductors. Photoluminescence quantum yields approaching unity in well-passivated samples add a thermodynamic version of the same statement: surface treatment of $MAPbI_3$ films can raise the internal photoluminescence quantum efficiency and quasi-Fermi-level splitting to within a few per cent of the radiative limit, rivalling the best GaAs absorbers[97], showing that the nonradiative channels can be made nearly silent without heroic bulk purity.

## 9.3. The counter-case

The counter-case attacks both conjuncts. On the electronic side, first-principles studies with careful finite-size, functional, and SOC treatment find that some abundant defects — notably interstitial and vacancy configurations of iodine in relevant charge states — do place levels deep enough to capture, and that the shallow-only picture partly reflected methodological artefacts of the early calculations[8]. On the kinetic side, the same school computes nonradiative capture coefficients — the quantity Shockley-Read-Hall recombination needs, which level positions alone do not supply. In its simplest single-trap form the nonradiative lifetime is

$$\tau_{SRH} \approx 1 / (C_n N_t) \tag{9}$$

with $C_n$, $C_p$ the electron and hole capture coefficients and $N_t$ the trap density: it is $C_n$ and $C_p$, not merely whether a level sits mid-gap, that this school's calculations target directly — and they find rates comparable to those of ordinary semiconductors: on this accounting halide perovskites are *not* exceptionally tolerant, and their long lifetimes must be purchased partly elsewhere (low formation of the specific killers, screening, healing). The recent critical synthesis of Mosquera-Lois *et al.*[9] audits the several proposed tolerance mechanisms — level positions, screening, polaronic protection, lattice softness reducing capture — and concludes that each is real in part and sufficient in none: "defect tolerance" as used in the literature is not one hypothesis but a family, with different members doing the work in different compounds and none yet quantitatively closed. We adopt that finding as this section's frame.

## 9.4. What the lattice adds: screening, capture, and disorder

Sections 4-7 supply three modifiers that the static level diagram omits. *Screening*: charged defects are subject to the same quasi-static dielectric response that dresses carriers (Section 6.3), which can reduce Coulomb capture cross sections. A further interpretation is that polaronic carrier dressing may modify the coupling between carriers and defect coordinates[10,11]. The resulting "polaron protection" is therefore a hypothesis for an additional dynamical contribution to defect tolerance, rather than an established universal mechanism. *Capture in a soft lattice*: nonradiative capture proceeds by multiphonon emission across a lattice-coordinate barrier, and the phonon energies, coupling strengths and anharmonicity of the accepting modes are all unusual here. Whether lattice softness suppresses or enhances capture remains an open question addressed by the first-principles calculations of Section 9.3. *Disorder*: polymorphous fluctuations can modulate defect levels themselves, so a “level” is better regarded as a distribution (Section 5.4), and static spectroscopy of defect states inherits the same averaging subtleties as the band structure. The coupled framework therefore provides plausible contributing mechanisms — screening, carrier dressing and lattice dynamics — but does not by itself determine defect formation or the net capture rate.

## 9.5. The mobile defect population and self-healing

The defects of this material move (Section 8.4)[60], and a mobile defect population is qualitatively different from silicon's frozen one: damage can anneal at room temperature, photo- and field-induced rearrangements are reversible on laboratory timescales, and the observed recovery of performance after stress — "self-healing" — has been framed as a distinct pillar of the material's forgiveness, raising the still-open question of whether defects here are healed, tolerated, or both[77]. Mechanistically, healing needs exactly what Sections 3–4 provide: low migration barriers in a soft open framework, shallow formation energetics, and a free-energy landscape in which the pristine configuration is recoverable. It is the kinetic face of the same softness whose electronic face is the polaron.

## 9.6. Adjudication

Our reading of the present evidence. (1) The band-edge-inversion mechanism is real and load bearing: it is why the typical defect is benign, and it survives its critics. (2) The strong claim — that no abundant defect captures efficiently — is not sustained; the capture-rate calculations of Zhang et al.8 stand unrefuted, and the observed lifetimes therefore require the cooperation of low killer-defect formation, screening and polaronic protection, and dynamic healing. (3) The question is no longer "is there defect tolerance" but a budget question — how many decades of nonradiative-rate suppression each mechanism contributes — and that budget has not been experimentally decomposed for even one compound. Its decomposition (systematic capture spectroscopy across the Cs/MA/FA and Cl/Br/I series, against the polaron and screening observables of Sections 6–7) is, in our view, the single most valuable experiment this field has not yet done. The framework proposed in Section 1.2 is therefore relevant to the defect story but is not yet sufficient to explain it — precisely the delimitation promised in Section 7.5.

# 10. Dimensional Reduction

Slicing the $[PbX_3]^-$ framework along low-index planes and intercalating organic spacers produces the layered Ruddlesden-Popper (RP) and Dion-Jacobson (DJ) families $A'_2A_{n-1}Pb_nX_{3n+1}$: natural multiple quantum wells whose inorganic thickness n is set by stoichiometry rather than epitaxy[18,33,78,18,94]. The essential point is not that dimensional reduction introduces an entirely new set of mechanisms, but that it changes their relative hierarchy. Quantum confinement, dielectric screening, lattice anharmonicity, carrier–lattice coupling, symmetry breaking and spin–orbit coupling is already present in the three-dimensional compounds. Reducing the dimensionality changes the spatial extent of the electronic states, the dielectric boundary conditions and the accessible lattice coordinates, thereby promoting some of these ingredients from corrections into dominant physics. Colloidal nanocrystals extend the same logic towards the zero-dimensional limit.

## 10.1. Quantum confinement and dielectric confinement

Two confinements act simultaneously. Quantum confinement is conventional: for n = 1, the carriers occupy a well approximately one octahedron thick, and the exciton becomes quasi-two-dimensional, with the familiar dimensional enhancement of binding. What becomes qualitatively more important in the layered compounds is dielectric confinement. The organic gallery has a small dielectric constant whereas the inorganic well has a large one (Section 6), so the electric field of the electron–hole pair extends into the poorly screening barrier. Image charges consequently reinforce the electron–hole attraction[18,79,82]. The measured consequence is an exciton binding energy of 370 meV in $(C_{10}H_{21}NH_3)_2PbI_4$, stable at room temperature, followed by systematic optical and magneto-optical studies[79,80,82–84]. Across the range of well thicknesses, Blancon et al.[81] found a systematic decrease in exciton binding energy from the hundreds-of-meV 2D limit towards the ~16 meV 3D limit. The dimensional trend therefore illustrates the central principle of this section: the exciton itself is not a new ingredient, but confinement and dielectric contrast move it from a relatively weakly bound 3D quasiparticle to the dominant room-temperature excitation of the layered compounds.

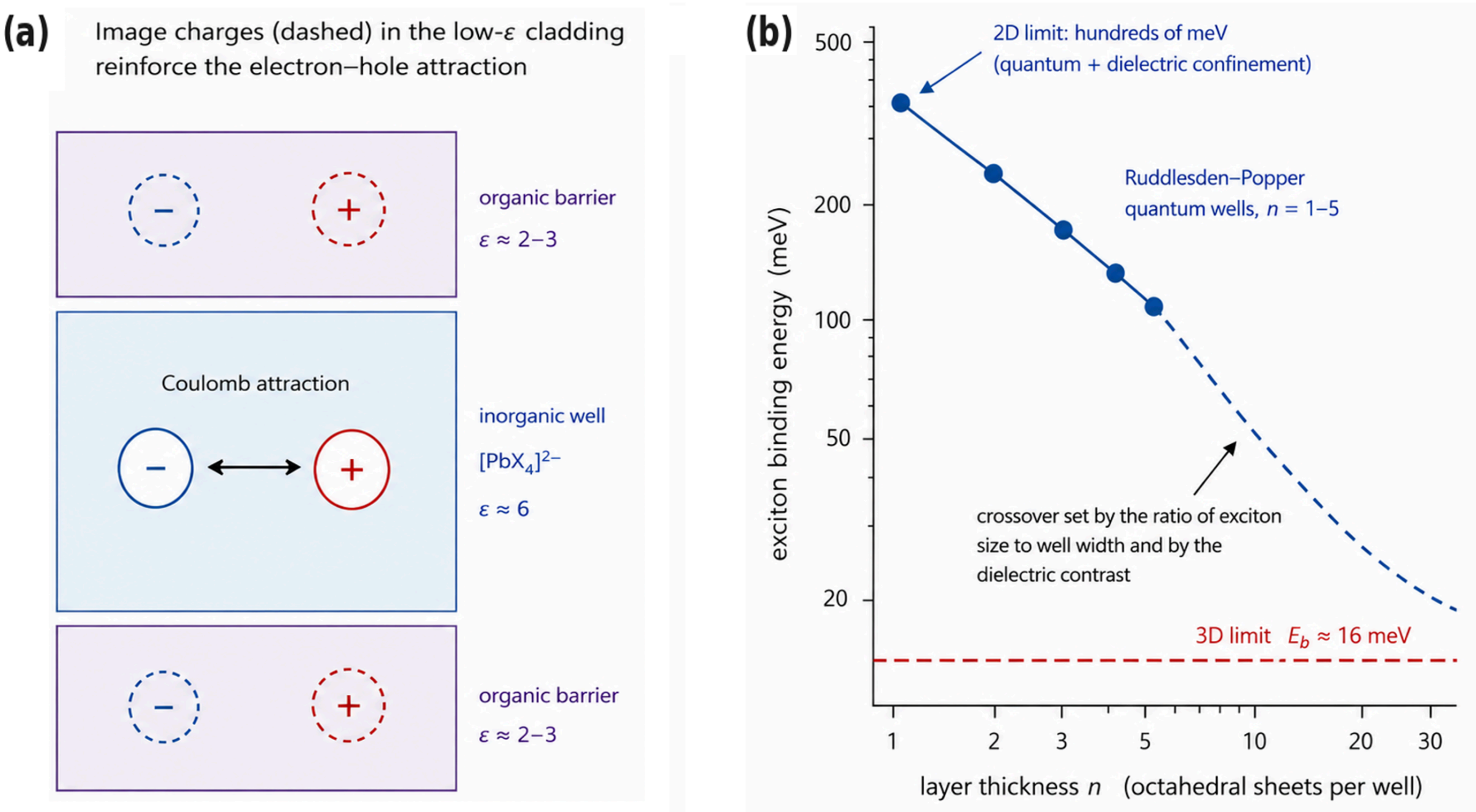

Panel (a): the dielectric contrast between well and barrier has no analogue in epitaxial III-V wells[18,79,82].
Panel (b) is schematic: for the measured scaling see Blancon *et al.*[81] and the Rydberg-series analysis[82,83].

**Fig. 20 | Dielectric confinement and exciton binding in 2D perovskites.** (a) Dielectric confinement, schematically: in a high-ε inorganic well clad by low-ε organic barriers, the image charges (dashed) of both carriers reinforce the electron-hole attraction, adding to quantum confinement an electrostatic binding channel with no analogue in epitaxial III-V wells[18,79,82]. (b) The resulting evolution of exciton binding with RP layer thickness *n*, schematically, interpolating between the measured 2D (hundreds of meV) and 3D (~16 meV) limits; measured scaling: Blancon *et al.*[81].

## 10.2. What the layers do to the Sections 4–9 machinery

Each element of the three-dimensional picture is modified, but not necessarily replaced. *Lattice*: in the three-dimensional compounds, low-energy octahedral rotations and correlated anharmonic fluctuations are properties of the connected $[PbX_3]^-$ framework (Section 4). Organic spacers interrupt this connectivity along the stacking direction. Dimensional reduction therefore changes the balance of lattice degrees of freedom and can make spacer-dependent structural distortions more prominent. *Electronic structure*: quantum confinement widens the band gap as n decreases, while the Pb-derived band edges and strong spin–orbit coupling remain the relevant electronic ingredients. The principal change is therefore not the origin of the electronic states but their spatial extent and symmetry. In genuinely non-centrosymmetric DJ and low-symmetry RP structures, Rashba physics can become an intrinsic structural property rather than a proposed consequence of fluctuating local symmetry breaking. *Screening and transport*: the dielectric response changes substantially because high-ε inorganic layers are separated by low-ε organic barriers. In-plane carrier motion can remain relatively mobile, whereas cross-plane transport becomes barrier-limited[85]. The same dielectric physics that modifies carrier interactions in 3D therefore becomes a primary control parameter in 2D. *Excited-state lattice coupling*:

the softer, lower-symmetry cages of many n = 1 compounds can support strong exciton-lattice coupling and self-localization, giving rise to broadband, heavily Stokes-shifted emission attributed to self-trapped excitons (Section 7.2). This places these materials towards the strong-coupling side of the polaron landscape of Figure 17, although the crossover from the 3D regime should not be regarded as a universal or abrupt boundary. Within this chemical family, dimensional reduction therefore provides a route from predominantly delocalized carrier–lattice coupling towards stronger exciton self-localization[18].

## 10.3. Nanocrystals and the fine structure surprise

Colloidal $CsPbX_3$ nanocrystals[86] extend the dimensional sequence towards zero dimensions and, with it, single-emitter spectroscopy: fast, spectrally narrow single-photon emission with reduced blinking at cryogenic temperature[87]. Here again, the central question is not which ingredients appear, but which terms become dominant as well all spatial dimensions approach the relevant excitonic length scales. In $CsPbX_3$ nanocrystals, Becker et al.[56] combined effective-mass and group-theoretical considerations, incorporating the j = 1/2 conduction-band character discussed in Section 5.2 and a Rashba term, and used size- and composition-dependent single-nanocrystal spectra to argue that the emitting triplet lies lowest. For $FAPbBr_3$ nanocrystals, however, magneto-optical fine-structure spectroscopy places a dark singlet below the bright triplet[88]. The bright-lowest ordering is therefore not a universal consequence of dimensional reduction, but a compound- and regime-dependent outcome of competing exchange, Rashba and shape contributions. This provides a clean example of the hierarchy principle: spin–orbit coupling, exchange interaction, structural asymmetry and confinement are all ingredients already present elsewhere in the review; zero-dimensional confinement changes their relative weights, making the ordering of individual exciton states directly observable.

locates a dark singlet below the bright triplet[88]. The bright-lowest ordering is therefore not a universal consequence of dimensional reduction, but a compound- and regime-dependent outcome of the competition among exchange, Rashba and shape terms. This provides a clean example of the hierarchy principle: spin–orbit coupling, exchange interaction, structural asymmetry and confinement are all ingredients already present elsewhere in the review; zero-dimensional confinement changes their relative weights enough that the ordering of individual exciton states becomes directly observable.

## 10.4. Status

Dimensional reduction therefore acts as a controlled modification of the framework developed in Sections 4–9. The principal ingredients do not disappear when the three-dimensional lattice is sliced: soft lattice dynamics, dielectric response, carrier–lattice coupling, strong spin–orbit coupling and structural symmetry all remain relevant. What changes is their hierarchy. In the 3D compounds, dielectric screening, polaronic dressing and correlated lattice dynamics modify an electronic structure that provides the most immediate description. In layered compounds, dielectric and quantum confinement elevate excitonic physics to the centre of the problem; in nanocrystals, confinement and fine-structure contributions can dominate individual optical transitions. Cross-plane transport, conversely, becomes a barrier-controlled process rather than a direct measure of intrinsic 3D band mobility. This shift in hierarchy is experimentally useful because dimensional reduction can partially separate mechanisms that are entangled in the bulk. The layer thickness n tunes confinement; the organic spacer tunes dielectric contrast and lattice connectivity; structural symmetry controls whether Rashba terms are conditional or intrinsic; and nanocrystal size tunes confinement and exciton fine structure. The layered and nanocrystalline families therefore provide complementary platforms for probing the same microscopic ingredients under different boundary conditions, rather than simply constituting lower-dimensional versions of the 3D compounds. The open questions follow directly from this hierarchy. The spacer dependence of anharmonic dynamics remains insufficiently mapped; the crossover between large-polaron and self-trapped exciton regimes requires a unified treatment; and exciton fine-structure ordering across composition, size and structural symmetry remains unresolved. The value of dimensional reduction is consequently not to add another catalogue of phenomena, but to reveal which of the existing ingredients becomes dominant as the hierarchy is changed.

# 11. Broken Symmetry and Emergent Phenomena

This section considers phenomena that depend on symmetry breaking — established, proposed or deliberately engineered — and on the spin and light–matter physics enabled by the strong spin–orbit coupling (SOC) discussed in Section 5. The central question is not simply whether symmetry is broken, but what form that breaking takes, on what length and time scales, and whether it produces a distinguishable physical response. This distinction is particularly important in lead-halide perovskites, where ionic motion, polymorphous disorder and strong SOC can generate local or dynamic symmetry breaking without necessarily producing a corresponding macroscopic order parameter. We therefore distinguish established structural responses from proposed microscopic interpretations and from phenomena in which symmetry breaking is used as a design variable.

## 11.1. The ferroelectricity question

The claim that $MAPbI_3$ is ferroelectric — polar, with switchable domains — circulated from the field's first years, attached to hopes of domain-assisted charge separation. The structural predicates are genuinely delicate: the tetragonal phase's space-group assignment (centrosymmetric I4/mcm versus polar I4cm) is difficult precisely because the candidate polar distortion is small and the molecular cations are disordered, and the measurement is confounded by the ionic conduction discussed in Section 8.4, which can mimic ferroelectric hysteresis in polarization loops. The affirmative case is put most strongly by Rakita *et al.*[89], who marshal pyroelectric, structural, and switching evidence for polar tetragonal $MAPbI_3$; ferroelasticity — switchable strain domains without net polarization — is separately and more robustly documented[90], with twin domains directly imaged. The evidence therefore supports a clear distinction between ferroelasticity and the stronger claim of switchable ferroelectric order. Ferroelasticity is established, whereas ferroelectricity in the strong, switchable sense remains unsettled. If a polar order parameter is present, its magnitude, temperature dependence and coupling to molecular-cation dynamics must be distinguished from local inversion breaking and ionic artefacts. The once-popular ferroelectric-domain explanation of photovoltaic performance is also not required by the intrinsic carrier, dielectric and lattice mechanisms developed in Sections 6–8, which provide alternative explanations[15]. The more durable physical possibility is therefore proximity to polar instability rather than established macroscopic ferroelectricity: a lattice close to a polar instability can sustain substantial anharmonic polar fluctuations without developing a static macroscopic polarization. This possibility connects the ferroelectricity debate to the dynamic symmetry breaking discussed below, but the two should not be treated as equivalent claims.

## 11.2. Spin physics

The ingredients — $j = 1/2$ band edges, giant SOC, conditional inversion breaking (Section 5.3) — make the family susceptible to spin-dependent responses[17]. In genuinely non-centrosymmetric structures, Rashba–Dresselhaus splitting can arise from static structural asymmetry (with the layered compounds of Section 10 supplying clean cases), whereas in centrosymmetric bulk materials the proposed dynamic-Rashba picture instead invokes fluctuating local inversion breaking. That last coupling gives spin relaxation in these materials an unusual character: the same slow polarization dynamics that dresses charge (Section 7) also dephases spin, so spin lifetimes interrogate the anharmonic lattice from an independent direction. At the single-object level, magneto-optics on individual nanocrystals reports Rashba-scale effects directly[91], and the bright-triplet fine structure of Section 10.3 is itself a spin-physics result — arguably the family's most concrete one[56,88]. These observations establish the relevance of spin-dependent interactions, but they do not by themselves establish a single microscopic origin for all Rashba-like responses across the material family. Chiral organic spacers provide a different route in which symmetry breaking is deliberately introduced. Incorporating a chiral cation can transmit structural chirality to the inorganic framework, enabling chiroptical and spin-selective responses in chiral 2D members. Whether the observed spin selectivity arises predominantly from structural chirality, interfacial electronic coupling or other contributions remains an important question for this developing area. We therefore flag chirality as an emerging direction rather than as an established universal consequence of organic–inorganic chirality transfer.

## 11.3. Light-matter hybrids

The large oscillator strengths and room-temperature-stable excitons of the 2D family (Section 10.1) provide the prerequisites for strong light–matter coupling, and exciton–polariton phenomena in perovskite microcavities are an active experimental frontier. Room-temperature polariton condensation has been reported in both three-dimensional and quasi-two-dimensional lead-halide perovskite microcavities, with condensation thresholds, coherence and nonlinear parametric scattering investigated95,96.

The relevance of these systems to the present section is not that polaritons constitute another independent mechanism, but that optical confinement reorganizes the same excitonic degrees of freedom into a hybrid light–matter state. The polariton is therefore a different dressed quasiparticle from the polaron of Section 7: the polaron reflects carrier–lattice coupling, whereas the polariton reflects coherent coupling between an excitonic transition and a photon mode.

The distinctive opportunity in lead-halide perovskites is that the excitonic component is embedded in an anharmonic, strongly polarizable and dynamically disordered lattice. Whether these lattice properties produce qualitatively distinct polariton behaviour, rather than simply modifying established cavity physics, remains an open question. We therefore do not attempt a comprehensive treatment of the polariton literature here; instead, we identify it as a natural extension of the coupled framework developed throughout the review.

## 11.4. Open questions

(1) A definitive, artefact-immune polar-order experiment for the tetragonal hybrids — the measurement protocol matters more than another sample. (2) Quantitative spin-relaxation spectroscopy across the Cs/MA/FA series, as an independent probe of the dynamic polar fluctuations. (3) Whether dynamical Rashba has any observable that cannot be mimicked by static local symmetry breaking in the polymorphous ensemble — a sharper theoretical question than is usually acknowledged. (4) The chiral and polaritonic programs, pending verified primary literature.

# 12. Conclusions and Outlook

A review that has insisted, section by section, on separating the measured from the inferred owes the reader a consolidated ledger. Table 3 collects, in one place, the verdict this review reaches on the field's most contested claims; Section 12.1 then closes with the seven problems we judge most consequential, ordered from the most concrete to the most conceptual, each stated with the experiment or calculation that would move it.

**Table 3.** Adjudication scoreboard for the field's most contested claims, as reached in the body of this review. “Status” reflects our reading of the balance of evidence at the time of writing, not a final resolution.

| Claim | Status | Basis |
|---|---|---|
| Defect tolerance — electronic basis (band-edge inversion pushes dangling-bond states out of the gap) | Supported | Founding calculations[7,75,76]; real and load-bearing (Sec. 9.6) |
| Defect tolerance — strong form (no abundant defect captures efficiently) | Not sustained | Capture-rate calculations[8] stand unrefuted; benignity is a multi-mechanism budget, not one fact (Sec. 9.6) |
| Ferroelasticity (switchable strain domains) | Established | Twin domains directly imaged[90] (Sec. 11.1) |
| Ferroelectricity (switchable polar domains, strong sense) | Unsettled | Pyroelectric/switching evidence[89], confounded by ionic conduction (Sec. 11.1) |
| Static bulk Rashba splitting (centrosymmetric phases) | Not supported | Ruled out by second-harmonic-generation centrosymmetry measurement[92] (Sec. 5.3) |
| Dynamical (fluctuation-driven) Rashba effect | Proposed, unresolved | Directly challenged by[92]; no discriminating observable yet identified (Sec. 5.3, Sec. 11.4) |
| $T^{-3/2}$ mobility law as an acoustic-phonon fingerprint | Rejected as a mechanism diagnostic | Linewidth data exclude acoustic-dominated scattering[13]; likely an emergent crossover slope (Sec. 8.3) |
| Unification hypothesis (Sec. 1.2): one soft-lattice origin for lattice, electronic, and dielectric anomalies | Supported for the lattice/electronic/dielectric sectors; not yet sufficient for defects | Defended in Sec. 7.5; defect chemistry adds an independent thermodynamic conjunct (Sec. 9.6) |

## 12.1. The ledger

**1. The nonradiative-rate budget (Section 9.6).** Decompose, for one compound, the decades of lifetime enhancement among killer-defect scarcity, dielectric screening, polaronic protection, and dynamic healing. Requires systematic capture spectroscopy (deep-level transient and photothermal methods, adapted to a mixed conductor) across the Cs/MA/FA and Cl/Br/I series, run against the same samples' polaron and screening observables. This is, in our judgement, the field's most valuable missing experiment, and the arbiter of how far the unification hypothesis of Section 1.2 extends.

**2. The two-dimensional polaron (Sections 4.6, 7.3).** The conjecture of Lanigan-Atkins *et al.*[4] — that the planar rotational correlations template anisotropic carrier dressing — is falsifiable with momentum-resolved probes of carriers (time-resolved diffuse scattering around a photoexcited population; anisotropy-resolved THz conductivity in detwinned crystals). A positive result would be a qualitatively new polaron; a negative one would usefully cap the correlation-transport coupling.

**3. The identity of the central peak (Section 4.6).** Polarization- and momentum-resolved quasielastic spectroscopy across the A-site and halide series, with self-consistent-phonon and MD theory run on the same compounds, to separate relaxational tilt hopping from overdamped soft-mode and cation-coupled channels. The answer calibrates every claim in Sections 4–8 that leans on "slow relaxational polarization."

**4. The transport representation problem (Section 8.3).** "Dressed carrier scattered by fluctuations" versus "carrier of a fluctuating band structure" must be separated by observables, not preferences: Hall factor and magnetotransport systematics, and broadband conductivity traced continuously across the solvation window from THz to optical. The $T^{-3/2}$ exponent should be retired as a mechanism diagnostic either way.

**5. The polar-order endgame (Section 11.1, 11.4).** One artefact-immune protocol — simultaneous structural, pyroelectric, and switching measurement with the ionic channel independently monitored — applied to the tetragonal hybrids, would end a decade of dispute. The physics payoff is the placement of the family relative to polar instability, which quantitatively anchors the anharmonic fluctuation amplitude of Section 4.

**6. Predictive anharmonicity (Section 4.6).** A theory that predicts, from composition, *which* anharmonic expression a compound selects[55] — central-peak weight, tilt-correlation dimensionality, transition topology (Figure 10) — would convert Sections 3–4 from description to design. The layered family (Section 10.2), where spacer chemistry tunes the tilt landscape, is the natural training set.

**7. The dressed-quasiparticle concept itself (Sections 5.5, 7).** When the lattice averaging time and the scattering time merge, what replaces the Bloch quasiparticle is not settled language anywhere in condensed matter; this family is the cleanest place to settle it. The prize is exportable: soft polar semiconductors are a class, not a compound, and the conceptual machinery built here — polymorphous electronic structure, dielectric solvation, anharmonically dressed transport — is the general theory of that class in draft form.

## 12.2. Outlook

The trajectory of this field has been from anomaly to mechanism: a device number that no conventional argument could produce (Section 2.3) forced a decade-long descent through the material's physics, and the descent bottomed out not in a new interaction or a new particle but in an old chemical fact — a soft, polar, lone-pair-bearing bond — whose consequences, followed honestly through lattice, dielectric, and electronic sectors, account for most of what startled the field and delimit what still does. If the perspective of this review is right, the lead-halide perovskites will be remembered less as a photovoltaic episode than as the system that taught condensed-matter physics how to think about crystals that are liquids in one sector and solids in another; and the seven problems above are the curriculum's unfinished chapters. We have marked throughout where we take positions; the reader is invited to hold us to the controls.

# Appendix A. Relativistic first-principles practice

The minimum honest workflow for the electronic structure of $APbX_3$, distilled from Sections 2.4.1 and 5. (1) *SOC is mandatory.* The Pb 6p spin-orbit splitting[38] is of order 1 eV; scalar-relativistic gaps that "agree" with experiment do so by cancellation against the exchange-correlation gap error, and every derived quantity — masses, alignments, defect levels — inherits uncancelled errors. (2) *Quasiparticle corrections belong on top of SOC, not instead of it.* The reference implementation is relativistic quasiparticle self-consistent GW[39]; modern practice, including the treatment of excitonic effects through BSE and the compromises of hybrid functionals, is surveyed by Filip and Leppert[58]. (3) *The geometry matters as much as the Hamiltonian.* Computing on the ideal cubic average is a category error (Section 3.5); polymorphous sampling — relaxing and averaging over locally distorted configurations — changes gaps at the tens-of-per-cent level and generates local spin physics invisible to the monomorphous cell[21]. (4) *Symmetry claims require symmetry care*: Rashba splittings computed in supercells with artificial polarity, or defect levels without SOC, populate part of the older literature and should be read accordingly[8].

# Appendix B. Molecular dynamics on anharmonic landscapes

Where Appendix A treats the electrons at fixed (ensembles of) geometry, the dynamical claims of Sections 4–8 rest on sampling the lattice motion itself. The essential points. (1) Harmonic and quasi-harmonic phonon calculations about the cubic geometry yield imaginary frequencies for the rotational branches — a *feature*, correctly diagnosing the multi-well landscape (Figure 13(a)), not a convergence failure. (2) Self-consistent phonon schemes restore real, temperature-dependent frequencies by renormalizing against thermal displacements, and are the right tool for renormalized dispersions and transition temperatures — but they linearize precisely the hopping kinetics that produce the central peak, for which direct molecular dynamics (ab initio or machine-learned potentials) is required. (3) The observables that connect simulation to the experiments of Section 4 are the displacement autocorrelations (central peak), the tilt-tilt spatial correlations (diffuse rods), and the instantaneous-gap trajectory (Section 5.4); the reference computations are Mayers *et al.*[19] and Schilcher *et al.*[20,42] for the coupled carrier-lattice problem and Zhao *et al.*[21] for the configurational statistics. (4) Finite-size caution: the quasi-two-dimensional correlations of Lanigan-Atkins *et al.*[4] span many cells; simulation cells that cannot contain them silently truncate the physics of Sections 4 and 7.3.

# Appendix C. The Fröhlich-Feynman polaron formalism

The formal spine of Sections 7–8. The Fröhlich Hamiltonian[63] couples a band carrier to LO phonons through the long-range polar interaction, with all material detail compressed into the dimensionless coupling constant

$$\alpha = (e^2/\hbar) \sqrt{m / (2\hbar\omega_{LO})}\, (1/\varepsilon_\infty - 1/\varepsilon_s) \tag{6}$$

built from the LO phonon frequency $\omega_{LO}$, the band mass $m$, and the difference of inverse dielectric constants that sets how much of the lattice polarization the carrier sees at that frequency (Section 6). Small-α perturbation theory gives the familiar mass renormalization

$$m_p = m\,(1 + \alpha/6 + \ldots) \tag{7}$$

large α gives self-trapping; the lead halides' α of roughly 1.7-3 sits in neither regime, and the workhorse is Feynman's all-coupling variational path integral[64], which replaces the phonon cloud by a fictitious mass on a spring and remains quantitatively accurate across the intermediate window. Mobility follows from the Feynman model via the path-integral response formalism, applied to $MAPbI_3$ with measured inputs by Frost[65]; the fully ab initio counterpart — the many-body electron-phonon vertex with multiphonon processes — is developed for this family by Schlipf, Poncé, and Giustino[40], and the general materials context is reviewed by Franchini *et al.*[16]. The formalism's standing assumption — harmonic phonons — is exactly what Section 4 qualifies; the open theoretical program (Section 12, problem 7) is the Fröhlich problem with a relaxational, correlated bath, for which the dynamic-disorder computations of Schilcher *et al.*[20,42] are the current state of the art.

# Acknowledgements

The authors acknowledge the use of generative artificial intelligence in the preparation of this review. Claude Opus 5 (Anthropic) and Grok 4.5 (xAI) were used to assist with the structural organization of the manuscript, the construction of the original schematic and conceptual figures, and English proofreading. The models were not used to generate scientific claims, to select or summarize literature without verification, or to produce bibliographic records; all references were checked against the primary or publisher record. The scientific content, the literature selection, the interpretation of the cited work, and all conclusions and adjudications presented here are the sole responsibility of the authors, who have reviewed and take full responsibility for every AI-assisted element of the manuscript.

# Conflict of Interest

The authors declare no conflict of interest.

# Data Availability Statement

Data sharing is not applicable to this article as no new data were created or analysed in this study.